\documentclass[iop]{emulateapj}

\usepackage{bm}

\shorttitle{Magnetic Braking Depenence on Coronal Temperature}
\shortauthors{Pantolmos \& Matt}

\begin{document}

\title{MAGNETIC BRAKING OF SUN-LIKE AND LOW-MASS STARS: DEPENDENCE ON
  CORONAL TEMPERATURE.}

\author{George Pantolmos and Sean P. Matt}
\affil{University of Exeter, Department of Physics \&
   Astronomy, Physics Bldg., Stocker Road, Exeter, EX4 4QL,
   UK;}

\begin{abstract}
Sun-like and low-mass stars possess high temperature coronae and lose
mass in the form of stellar winds, driven by thermal pressure and
complex magnetohydrodynamic processes. These magnetized outflows
probably do not significantly affect the star's structural evolution
on the Main Sequence, but they brake the stellar rotation by removing
angular momentum,  a mechanism known as magnetic braking. Previous
studies have shown how the braking torque depends on magnetic field
strength and geometry, stellar mass and radius, mass-loss rate, and
the rotation rate of the star, assuming a fixed coronal
temperature. For this study we explore how different coronal
temperatures can influence the stellar torque. We employ 2.5D,
axisymmetric, magnetohydrodynamic simulations, computed with the PLUTO
code, to obtain steady-state wind solutions from rotating stars with
dipolar magnetic fields. Our parameter study includes 30 simulations
with variations in coronal temperature and surface-magnetic-field
strength. We consider a Parker-like (i.e. thermal-pressure-driven)
wind, and therefore coronal temperature is the key parameter
determining the velocity and acceleration profile of the flow. Since
the mass loss rates for these types of stars are not well constrained,
we determine how torque scales for a vast range of stellar mass loss
rates. Hotter winds lead to a faster acceleration, and we show that
(for a given magnetic field strength and mass-loss rate) a hotter
outflow leads to a weaker torque on the star. We derive new predictive
torque formulae for each temperature, which quantifies this effect
over a range of possible wind acceleration profiles. 
\end{abstract}


\section{Introduction} \label{sec_introduction}

Stellar winds are a very common phenomenon in our universe. For
Sun-like and low-mass stars ($M_* \lesssim 1.3M_{\odot}$), such
outflows are usually in the form of coronal winds
\citep{Parker:1958aa,Parker:1963aa}, due to their origin in the
several MK stellar hot coronae. Although the effect of coronal winds
on stellar mass during a star's Main-Sequence (MS) life is relatively
small, they can influence the environment of surrounding planets
\citep[e.g.][]{Luftinger:2015aa}, and have an enormous impact on
stellar rotation by exerting a spin-down torque on the stellar surface
\citep[e.g.][]{Schatzman:1962aa,Weber:1967aa}. Hence, over the years
the angular momentum (or rotational) evolution of cool stars has been
the subject of very intensive studies \citep[for a review
see][]{Bouvier:2014aa}.

The spin-down of MS cool stars was established observationally from
early studies \citep{Kraft:1967aa, Skumanich:1972aa} that showed the
rotation periods of these types of stars to increase as the stellar age
advances. The current picture of the rotational evolution of cool
stars is more complicated, and observations  
\citep[e.g.][]{Barnes:2003aa,Barnes:2010aa,Irwin:2009aa,Meibom:2011aa,Meibom:2015aa}
show that stellar rotation depends on both the mass and age. In
addition, the observed trends between magnetic activity  
(or coronal X-ray emission) and stellar rotation
\citep[e.g.][]{Pizzolato:2003aa,Wright:2011aa}, and the observed
evolution of stellar magnetic properties
\citep[e.g.][]{Vidotto:2014ab,See:2015aa} suggest that solar- and
late-type stars lose mass and angular momentum in the form of
magnetized outflows. 

Coronal-wind modeling has a long history in the literature, with the
use of analytic theory
\citep[e.g.][]{Parker:1958aa,Weber:1967aa,Mestel:1968aa,Heinemann:1978aa,Low:1986aa},
or iterative methods/numerical simulations 
\citep[e.g.][]{Pneuman:1971aa,Sakurai:1985aa,Washimi:1993aa,Keppens:2000aa,Cohen:2007aa,Vidotto:2009ab}. 
The main source for understanding the nature, the properties and the
dynamics of coronal winds comes from direct observations of the solar
wind. The solar corona expands into the interplanetary space in the
form of a supersonic, magnetized wind that evolves during a solar
cycle. Near the solar minimum the solar wind is bimodal with a fast,
tenuous, and steady, component emanating from large polar coronal
holes and a slower, denser and filamentary component emerging from the
top of the helmet streamers originated at the magnetic activity belt
\citep[e.g.][]{McComas:2007aa,McComas:2008aa}. During the solar
maximum the solar wind becomes more variable and is more dominated by
the slow wind at all latidtudes
\citep[e.g.][]{McComas:2003aa,McComas:2007aa}. The solar wind is a
direct consequence of the hot solar corona (with $T>10^6K$) and thus
the solar-plasma acceleration (for both the fast and slow solar wind)
is connected to the coronal heating problem
\citep[e.g.][]{De-Moortel:2015aa}. The physical mechanisms responsible
for the solar-corona heating are still in debate, but they all require
magnetic fields as a key ingredient \citep[see,
e.g.][]{Aschwanden:2005aa,Klimchuk:2015aa,Velli:2015aa}. The solar
magnetic field (a product of the solar dynamo that operates within the
convection zone) threads the solar photosphere, expands throughout the
solar atmosphere and eventually connects with and energizes the solar
wind. The recent advances in solar-wind theory include wave
dissipation (via turbulence) and magnetic reconnection as heat sources
for the expanding outer solar atmosphere \citep[see,
e.g.][]{Ofman:2010aa,Cranmer:2012aa,Cranmer:2015aa,Hansteen:2012aa}. Scaling-law
models
\citep[e.g.][]{Wang:1991aa,Fisk:2003aa,Schwadron:2003aa,Schwadron:2008aa}
reproduce part of the observed characteristics of the solar wind,
although, that approach does not treat the coronal heating/solar-wind
acceleration problem in a self-consistent way \citep[see,
e.g.][]{Hansteen:2012aa}. A conclusive answer on what heats the solar
corona and what are the physical processes that drive the solar wind
does not exist. X-ray observations have revealed the existence of hot
outer atmospheres in every low-mass star
\citep[e.g.][]{Wright:2011aa}. However, it is still not clear how
coronal heating should vary among late-type stars with varying masses
and rotation rates, and what does this indicate for the observed
trends in X-ray emission \citep[see, e.g.][]{Testa:2015aa}. Therefore,
it is still an open question of how to apply our knowledge of the
solar coronal heating and wind acceleration to other stars. The
present work is concerned with characterizing the global torques on
stars and how they scale for a variety of stellar properties, while
solutions to the coronal-heating problem remain
uncertain. Consequently, in this work, we treat many of the coronal
processes as "free parameters'', including the wind mass loss rates
and  wind acceleration profiles, which show how the uncertainties in
our understanding of stellar coronae will influence our ability to
predict angular momentum loss. 

In the framework of stellar-torque theory, early works
\citep[e.g.][]{Schatzman:1962aa,Mestel:1984aa,Mestel:1987aa,Kawaler:1988aa}
have provided analytic prescriptions for the magnetic braking of cool
stars, and some more recent works compute the stellar 
angular momentum losses self-consistently, via multidimensional
numerical simulations. For example, studies have quantified how the
magnetic braking scales with various stellar parameters
\citep[e.g.][]{Matt:2008aa,Matt:2012ab,Cohen:2014aa}, and others have
showed how stellar angular momentum losses depends on different
magnetic field geomtries
\citep[e.g.][]{Garraffo:2015ab,Garraffo:2016aa,Reville:2015ab,Finley:2017aa}. With 
the new advances in Zeeman-Doppler Imaging
\citep[e.g.][]{Donati:1997aa,Donati:2009aa}, observers can now extract
stellar-surface magnetic field maps that can be used in order to
reconstruct the stellar field near the star. Some studies 
\citep[e.g.][]{Vidotto:2014aa,Alvarado-Gomez:2016ab,Reville:2016aa},
have used such maps in their wind simulations, in order to provide
trends for stellar torques based on realsitic magnetic
fields. In general, accurate stellar-torque predictions are one of the
critical ingredients for rotational evolution models
\citep[e.g.][See et
al. submitted]{Reiners:2012aa,Gallet:2013aa,Gallet:2015aa,Johnstone:2015ad,Matt:2015aa,Amard:2016aa}.

Coronal temperatures among MS cool stars significantly vary
\citep[e.g.][]{Johnstone:2015aa}. However, there has not yet been a
systematic study of magnetic braking that investigates the key
parameters (i.e. stellar coronal temperature and polytropic index),
that affect the wind driving (or flow acceleration and
velocity). The objective of this study is to quantify the influence of
different flow temperatures on stellar torques. We adopt the approach
introduced in \citet{Matt:2008aa}. In particular, \citet{Matt:2008aa}
found that the effective magnetic lever arm (or Alfv\'{e}n radius),
that determines the efficiency of the braking torque, is a power law
in a parmeter $\Upsilon$ (i.e. wind magnetization), that depends on
the stellar mass, radius, mass-loss rate, and magnetic field
strength. Studies on massive, hot stars \citep[e.g. type O stars,
see][]{ud-Doula:2009aa}, have found similar scalings between the
stellar paramters and angular momentum losses, with the main
difference being that the wind-driving mechanism is fundamentally
different \citep[e.g.][]{Lamers:1999aa,Owocki:2009aa}. Following
\citet{Matt:2008aa}, a series of studies
\citep[][]{Matt:2012ab,Reville:2015ab,Reville:2016aa,Finley:2017aa}, expanded the 
previous torque formulation in braking laws that include the depedence
of the braking torque on the stellar spin rate and different magnetic
field geometries. All these studies
\citep{Matt:2008aa,Matt:2012ab,Reville:2015ab,Reville:2016aa,Finley:2017aa} 
used polytropic, Parker wind models 
\citep[e.g.][]{Parker:1963aa,Keppens:1999aa,Lamers:1999aa}, modified
by rotation and magnetic fields. However, they kept fixed the flow
thermodynamics (i.e. coronal temperature and polytropic index), that
determine the wind velocity and acceleration. 

The purpose of this paper is to examine, and quantify how variations
in coronal temperature (one of the key parameters that influence the wind
accelaration) will affect the stellar angular momentum loss, employing
2.5D, ideal MHD, and axisymmetric, simulations. In the following
section (\S \ref{sec_SWtheory}), we provide a brief theoretical
discussion on the concept of angular momentum loss due to stellar
outflows. In section \ref{sec_WindSols}, we discuss how our numerical
setup is suited to study a wide range of wind acceleration profiles,
and describe our parameter space. In section \ref{sec_GlobProp} we
focus on the results of this study, and we show braking laws for
different temperartures. In section \ref{sec_BrakingLaws}, two new
torque formulae that are independent of the flow temperature are
proposed, and finally in section \ref{sec_Conclusions} the main
conclusions of this paper are summarized. In Appendices
\ref{Appendix_SteadyState}, and \ref{Appendix_Accuracy} we discuss
some numerical issues in our simulations. Appendix
\ref{Appendix_Kq_q_vs_CsVesc} provides an empirical approach to
predict stellar torques for any temperature. Finally, Appendix
\ref{Appendix_Sims} contains plots of the complete simulation grid for
this parameter study.

\section{Magnetized outflows and efficiency of angular momentum loss} \label{sec_SWtheory}

In general, the total angular momentum rate carried away from a star
in a stellar wind can be written as
\begin{eqnarray}
\label{eq_2Dtorque}
\tau_w=\dot{M}_w \Omega_* <R_A>^{2},
\end{eqnarray}
where $\dot{M}_w$ is the integrated stellar mass loss rate due to the
wind, $\Omega_*$ is the stellar rotation rate and $<R_A>^2$ is the
square of a characteristic length scale in the wind. Using a
mechanical analogy, $<R_A>$ can be thought of as a "lever arm length"
that determines the efficiency of the torque on the star exerted by
the plasma efflux. Generically, this efficiency of the angular
momentum loss can be expressed as the ratio of this lever arm length
to the stellar radius, $R_*$, 
\begin{eqnarray}
\label{eq_rA}
{<R_A> \over R_*} \equiv \left( {\tau_w} \over {\dot{M}_w \Omega_*
  R_*^2} \right)^{1/2}.
\end{eqnarray}
The precise value for the lengthscale $<R_A>$ depends on the detailed
(and multi-dimensional) physics of the wind. As an example, a
spherically symmetric, inviscid, hydrodynamical wind would simply
carry away the specific angular momentum it has from the stellar
surface. Thus the star is subjected to an agular momentum loss that
gives  $<R_A>/R_* = (2/3)^{1/2}$\citep[e.g.][]{Mestel:1968aa}, which
deviates from unity because the torque depends on the distance from
the rotation axis (i.e. cylindrical $\varpi = r \sin \theta$), not the
spherical radius $r$.

In a magnetized wind, Lorentz forces transmit angular momentum from
the star to the wind, even after it has left the stellar surface,
which can significantly increase the effieciency of angular momentum
loss. \citet[][see also \citet{Schatzman:1962aa}]{Weber:1967aa},
showed that for a one-dimensional, magnetized flow along the stellar
equator, under the assumption of steady-state, ideal MHD, this radius
equals the radial Alfv\'{e}n radius, defined as the radial distance
where the wind speed equals the local Alfv\'{e}n speed (considering
only the radial components of the velocity and magnetic field). In a
two or three-dimensional, ideal MHD flow, the value of $<R_A>^2$ is
the mass-loss-weighted average of the square of the poloidal
Alfv\'{e}n (cylindrical) radius \citep{Washimi:1993aa}.

In our simulations, $\Omega_*$ and $R_*$ are specified as input
parameters, and we directly compute the resulting values
of $\tau_w$ and  $\dot{M}_w$ in the wind solutions (see below). Thus,
following \citet{Matt:2008aa},
\citep[and][]{Matt:2012ab,Reville:2015ab,Reville:2016aa,Finley:2017aa}, we compute 
the value of $<R_A>/R_*$ using equation (\ref{eq_rA}) and refer to this
throughout as the "torque-averaged Alfv\'{e}n radius'' or "effective
Alfv\'{e}n radius". Note that, defining $R_A$ in this way does not
depend on any assumptions about the physics of the angular momentum
transfer (e.g., it does not require a steady-state, nor assume ideal
MHD conditions); the value $(<R_A>/R_*)^2$ simply represents a
dimensionless torque. Also, the scaling laws we derive below for
predicting $<R_A>$ are, by definition, the appropriate lenghtscale to
use in equation (\ref{eq_2Dtorque}) for computing the global torque.

\section{STELLAR WIND SOLUTIONS} \label{sec_WindSols}

\subsection{Numerical Setup} \label{subsec_NumSet}

This study employs ideal MHD and axisymmetric simulations, using the
PLUTO code \citep{Mignone:2007aa} in a 2.5D computational grid (i.e. 2
spatial coordinates with three vector components), in order to obtain
steady-state (or quasi-steady-state) stellar wind solutions. PLUTO
numerically solves the following set of ideal MHD conservation laws:
\begin{eqnarray}
\label{mass_cont}
\partial_{t}\rho + \nabla\cdot\rho \bm{\upsilon} = 0,
\end{eqnarray}
\begin{eqnarray}
\label{mom_eqn}
\partial_{t} \bm{m} + \nabla \cdot (\bm{m \upsilon} -
  \bm{B B} + \bm{\mathrm{I}} p_{tot}) = \rho \bm{g},
\end{eqnarray}
\begin{eqnarray}
\label{energ_eqn}
\partial_{t} E + \nabla \cdot [(E + p_{tot}) \bm{\upsilon}
  -\bm{B} (\bm{v} \cdot \bm{B})] = 
 \bm{m} \cdot \bm{g}, 
\end{eqnarray}
\begin{eqnarray}
\label{ind_eqn}
\partial_{t} \bm{B} + \nabla \cdot (\bm{\upsilon B - B \upsilon})
  = 0, 
\end{eqnarray}
where $\partial_{t} \equiv \partial/\partial t$ denotes the time
derivative operator, and $\bm{\mathrm{I}}$ is the identity
matrix. The mass density is denoted by $\rho$, $p_{tot} = p
+\bm{B}^2/2$ is the total pressure, composed of the thermal
pressure, $p$, and the magnetic pressure \footnote{In the PLUTO code
  the magnetic field is defined with a factor of $1/\sqrt{4 \pi}$
  included.}, $\bm{B}^2/2$. The velocity field is
$\bm{\upsilon}$, $\bm{m}=\rho \bm{\upsilon}$ is the
momentum denisty, $\bm{B}$ is the magnetic field, and
$\bm{g}=-(GM_*/r^2) \widehat{r}$ represents the gravitational
acceleration, where $G$ is Newton's gravitational constant, $M_*$ is
the stellar mass, $r$ is the distance to center of the star, and
$\widehat{r}$ stands for the radial unit vector. The total energy
density is $E = \rho e + \bm{m}^2/(2 \rho) +
\bm{B}^2/2$, where $e$ is the specific internal
energy. Finally, we adopt an equation of state for ideal gases, $\rho
e = p/(\gamma - 1)$, where $\gamma$ is the adiabatic exponent. 

\begin{figure}
\plotone{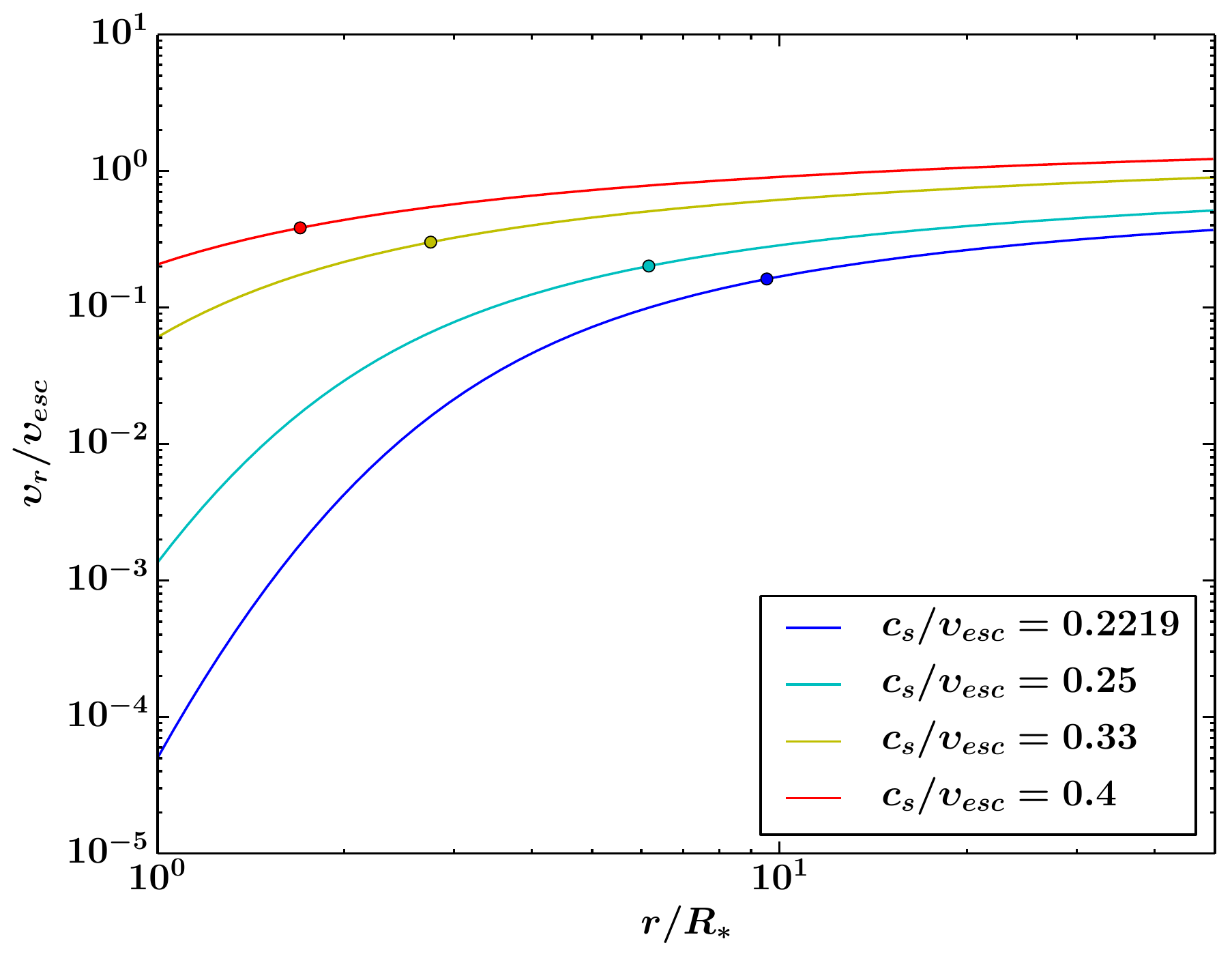}
\caption{Flow velocity versus radial distance for four different
  temperatures, here parameterized by the ratio of the adiabatic sound
  speed to the escape speed from the star, for one-dimensional,
  hydrodynamic winds from non-rotating stars. The above profiles are
  also used as the initial velocity wind profile in our
  simulations. The circles correspond to the radial distance at which
  the flow becomes supersonic. Each temperature produces a unique wind
  acceleration profile and hotter winds always exhibit higher base and
  terminal velocities than cooler winds.} 
\label{HDwind_log_Vr}
\end{figure}

We use a second-order piecewise linear reconstruction of all the
primitive variables ($\rho, \bm{\upsilon}, p, \bm{B}$) with
minmod limiter, and HLL Riemann solver \citep[e.g.][]{Toro:2009aa}
to compute the fluxes in equations (\ref{mass_cont}) -
(\ref{ind_eqn}). The induction equation (eq. \ref{ind_eqn}) is solved
with the constrained transport (CT) method \citep{Balsara:1999aa} in
order to ensure that the divergence-free condition for the magnetic
field will be maintained in our domain. The computational gird has
spherical geometry for the spatial coordinates, and covers $r \in [1,
50]R_*$, where $R_*$ is the stellar radius, and $\theta \in [0,\pi]$, 
with a total of $256 \times 512$ zones. A strecthed grid is
constructed  along $\widehat{r}$ direction. The first grid zone at the
stellar surface (i.e. inner boundary where $r/R_* = 1$) has size
$\Delta r= 5 \times 10^{-3}R_{*}$ but increases with $r$ such that 256
points reach $50R_*$ (i.e. outer boundary), with the last grid cell
having size $\Delta r= 1.015R_{*}$. The grid is uniform along the
$\widehat{\theta}$ direction.

We initialize the whole computational domain with a dipole field,
for which the radial, and polar componetnts are given by
\begin{eqnarray}
\label{eq_Br}
B_r = 2B_* \left(R_* \over r \right)^3 \cos\theta,
\end{eqnarray}
\begin{eqnarray}
\label{eq_Btheta}
B_{\theta}  = B_* \left(R_* \over r \right)^3 \sin\theta,
\end{eqnarray}
where $B_*$ is the equatorial surface field strength. We treat the
magnetic field using the the "background field splitting" approach
\citep{Powell1999284}, which sets the dipole field as a
time-independent compenent, and the code calculates the deviation from
the initial field. This method provides better numerical accuracy
in the treatment of the magnetic field, especially where strong
gradients in the magnetic field might otherwise lead to significant
numerical diffusion.  

We also initialize our grid with a 1D, polytropic, Parker's wind
solution shown in Figure \ref{HDwind_log_Vr}, and we set the density
and the thermal pressure based on the mass continuity equation and the
polytropic relation ($p_{th} \propto \rho^{\gamma}$),
respectively. Further details can be found in the following subsection
(\S \ref{subsec_Params}).

For both boundary zones of the $\theta$ coordinate we use an
"axisymmetric" type of boundary condition, which symmetrizes all the
variables across the borders and flips the signs of the $\phi$ and
normal components of the vector fields. The outer boundary condition
of $r$ coordinate is set to be "outflow", which sets the gradient of
each variable to be zero across the boundary. When the code stars to
evolve equations (\ref{mass_cont}) - (\ref{ind_eqn}) in time, the
initial state is blown outwards and the steady-state solution, we are
interested in, depends only on the inner boundary conditions. Since
our wind solutions only depend on the inner boundary, that represents
the stellar surface, for these ghost zones we use a more sophisticated
boundary condition. We keep fixed at the stellar boundary the values
for the thermal pressure and density computed from the one-dimensional
polytropic Parker's wind, we used to initialize our grid. This
boundary condition corresponds to a stellar atmosphere in which its
density and temperature do not vary in time and exhibit a temperature
profile such  that $T \propto \rho^{\gamma-1}$. Moreover, this
condition ensures that the temperature of the flow does not exhibit a
depedence on $\theta$ at the stellar boundary. The boundary condition
for the poloidal magnetic field is forced  to maintain the initial dipole
state, since the flow is sub-alfv\'{e}nic and magnetic pressure
dominates over the thermal and wind's hydrodynamic pressure. For the
toroidal magnetic field, we linearly extrapolate the toroidal field
values calculated in the computational domain into the ghost
zones. For the poloidal velocity, we also linearly extrapolate the
computed value of the poloidal velocity into the ghost zones, in order
to have a flow velocity that increases monotonically with radius
inside the ghost zones. In a steady-state  and axisymmetric flow, the
torroidal component of the electric field should be zero
\citep[e.g.][]{Lovelace:1986aa,Zanni:2009aa} and thus we force  the
poloidal component of the velocity and magnetic field to be parallel
to each other. The rotation is enforced only in the stellar boundary,
which we accomplish by setting the boundary condition for the toroidal
component of the velocity given by the equation: 
\begin{eqnarray}
\label{eq_Vphi}
\upsilon_{\phi} = \Omega_{*} r \sin\theta + {\upsilon_p \over B_p } B_{\phi},
\end{eqnarray}
in order to satisfy the $\bm{E}=0$ condition in a frame
rotating with the star \citep{Zanni:2009aa}. In equation
(\ref{eq_Vphi}), $r$ is the spherical radius and the subscripts $p$
and $\phi$ stand for the poloidal and torroidal components
respectively, of the velocity and magnetic field. 

Each simulation is stopped when the solution converges to a
steady-state. Some of the obtained numerical solutions are periodic,
and we discuss the steadiness, and the peculiarity of these
simulations in Appendix \ref{Appendix_SteadyState}.  We further
examine the correctness of each wind solution by checking how well the
five constant of motion are conserved along the flow streamlines
\citep[e.g.][]{Keppens:2000aa}. The numerical accuracy of our
simulations is discussed in more detail in Appendix
\ref{Appendix_Accuracy}.

\subsection{Parameters of the Study} \label{subsec_Params}

\begin{deluxetable*}{ccccc}
\tablewidth{0pt}
\tablecaption{Coronal Temperatures of the Parameter Study for
  Different Stellar Properties \label{tab_coronalT}}
\tablehead{
\colhead{} &
\colhead{Temperature ($\textrm{MK}^{\circ}$)} &
\colhead{Temperature ($\textrm{MK}^{\circ}$)} &
\colhead{Temperature ($\textrm{MK}^{\circ}$)} &
\colhead{Temperature ($\textrm{MK}^{\circ}$)} \\ 
\colhead{$c_s/\upsilon_{esc}$} & 
\colhead{$M_* = 1M_{\odot}$} &
\colhead{$M_* = 0.7M_{\odot}$} &
\colhead{$M_* = 0.5M_{\odot}$} &
\colhead{$M_* = 0.2M_{\odot}$} \\ &
\colhead{$R_* = 1R_{\odot}$} &
\colhead{$R_* = 0.65R_{\odot}$} &
\colhead{$R_* = 0.47R_{\odot}$} & 
\colhead{$R_* = 0.22R_{\odot}$}
}

\startdata
0.2219 & 1.30 & 1.40 & 1.39 & 1.20\\
0.25 & 1.65 & 1.77 & 1.77 & 1.52 \\
0.33 & 2.88 & 3.09 & 3.08 & 2.66 \\
0.4 & 4.23 & 4.53 & 4.53 & 3.90\\
\enddata

\end{deluxetable*}

\begin{deluxetable*}{cccccccc}
\tablewidth{0pt}
\tablecaption{Simulation Input Parameters and Resulting Global Wind
  Properties \label{tab_results_a}}
\tablehead{
\colhead{Case} &
\colhead{$c_s/\upsilon_{esc}$} &
\colhead{$\upsilon_A/\upsilon_{esc}$} &
\colhead{$\Upsilon$} &
\colhead{$<R_A>/R_*$} &
\colhead{$\Upsilon_{open}$} &
\colhead{$\Phi_{open}/\Phi_*$} &
\colhead{$\bar{\mathrm{V}}_{R_A}/\upsilon_{esc}$}
}

\startdata
1 & 0.2219 & 0.0151 & 2.90 & 3.62 & 283 & 0.787 & 0.0567\\
2 & 0.2219 & 0.0301 & 11.9 & 5.52 & 1020 & 0.737 & 0.128\\
3 & 0.2219 & 0.0452 & 27.7 & 6.23 & 1470 & 0.581 & 0.146\\
4 & 0.2219 & 0.0753 & 79.9 & 7.27 & 2330 & 0.430 & 0.17\\
5 & 0.2219 & 0.105 & 157 & 8.07 & 3170 & 0.358 & 0.187\\
6 & 0.2219 & 0.301 & 1240 & 11.8 & 9810 & 0.224 & 0.264\\
7 & 0.2219 & 0.627 & 5980 & 16.5 & 25600 & 0.165 & 0.335\\
8 & 0.2219 & 0.953 & 15000 & 20.2 & 44300 & 0.137 & 0.374\\
9 & 0.2219 & 1.51 & 41200 & 25.3 & 81300 & 0.112 & 0.415\\
10 & 0.25 & 0.21 & 33.2 & 4.71 & 1170 & 0.473 & 0.206\\
11 & 0.25 & 0.301 & 69.1 & 5.47 & 1820 & 0.409 & 0.236\\
12 & 0.25 & 0.627 & 335 & 7.83 & 5070 & 0.309 & 0.312\\
13 & 0.25 & 0.953 & 899 & 9.83 & 9460 & 0.258 & 0.361\\
14 & 0.25 & 1.51 & 2720 & 12.7 & 18600 & 0.208 & 0.413\\
15 & 0.25 & 2.5 & 8990 & 16.8 & 38200 & 0.164 & 0.465\\
16 & 0.25 & 4.14 & 29100 & 22.0 & 75700 & 0.128 & 0.512\\
17 & 0.33 & 0.953 & 16.7 & 3.27 & 1300 & 0.704 & 0.453\\
18 & 0.33 & 2.5 & 173 & 5.79 & 5470 & 0.448 & 0.609\\
19 & 0.33 & 3.01 & 275 & 6.47 & 7180 & 0.406 & 0.639\\
20 & 0.33 & 4.14 & 612 & 7.86 & 11400 & 0.344 & 0.683\\
21 & 0.33 & 6.2 & 1650 & 10.1 & 20700 & 0.282 & 0.736\\
22 & 0.33 & 11 & 6630 & 14.3 & 45600 & 0.209 & 0.802\\
23 & 0.33 & 17.5 & 20500 & 18.6 & 85000 & 0.162 & 0.845\\
24 & 0.4 & 4.14 & 194 & 5.68 & 7900 & 0.507 & 0.904\\
25 & 0.4 & 6.2 & 505 & 7.27 & 13800 & 0.416 & 0.969\\
26 & 0.4 & 8.6 & 1090 & 8.76 & 20800 & 0.348 & 1.01\\
27 & 0.4 & 11 & 1960 & 10.2 & 28800 & 0.305 & 1.04\\
28 & 0.4 & 17.5 & 5890 & 13.0 & 50800 & 0.234 & 1.10\\
29 & 0.4 & 26 & 13700 & 16.4 & 90400 & 0.204 & 1.14\\
30 & 0.4 & 50 & 62700 & 22.7 & 193000 & 0.140 & 1.21\\
\enddata

\end{deluxetable*}

\begin{figure}
\plotone{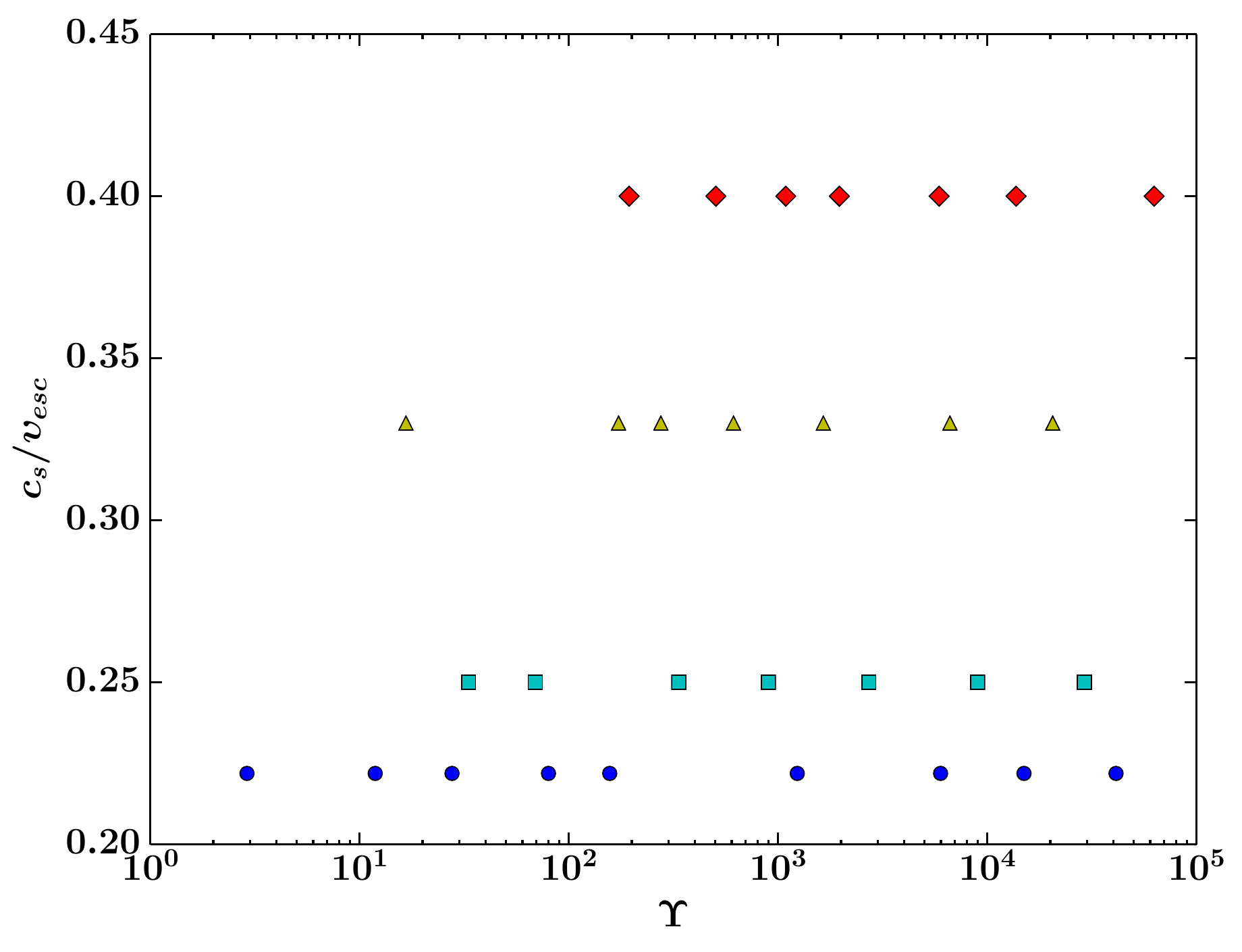}
\caption{Parameter space for the 30 simulations in this
  study. The vertical axis shows parameter $c_s/\upsilon_{esc}$, which
  controls the flow temperarture. The horizontal axis shows the
  paremeter $\Upsilon$, which is the wind magnetization (see eq
  \ref{eq_Upsilon}), and is associated with the average,
  stellar-surface magnetic field strength. Circles (blue), squares
  (cyan), triangles (yellow), and diamonds (red) correspond to
  simulations with $c_s/\upsilon_{esc}=$ 0..219, 0.25, 0.33, and 0.4,
  respectively. Every symbol represents a single case, for which we
  have a steady-state, wind solution.}
\label{ParamSpac}
\end{figure}

\begin{figure}
\plotone{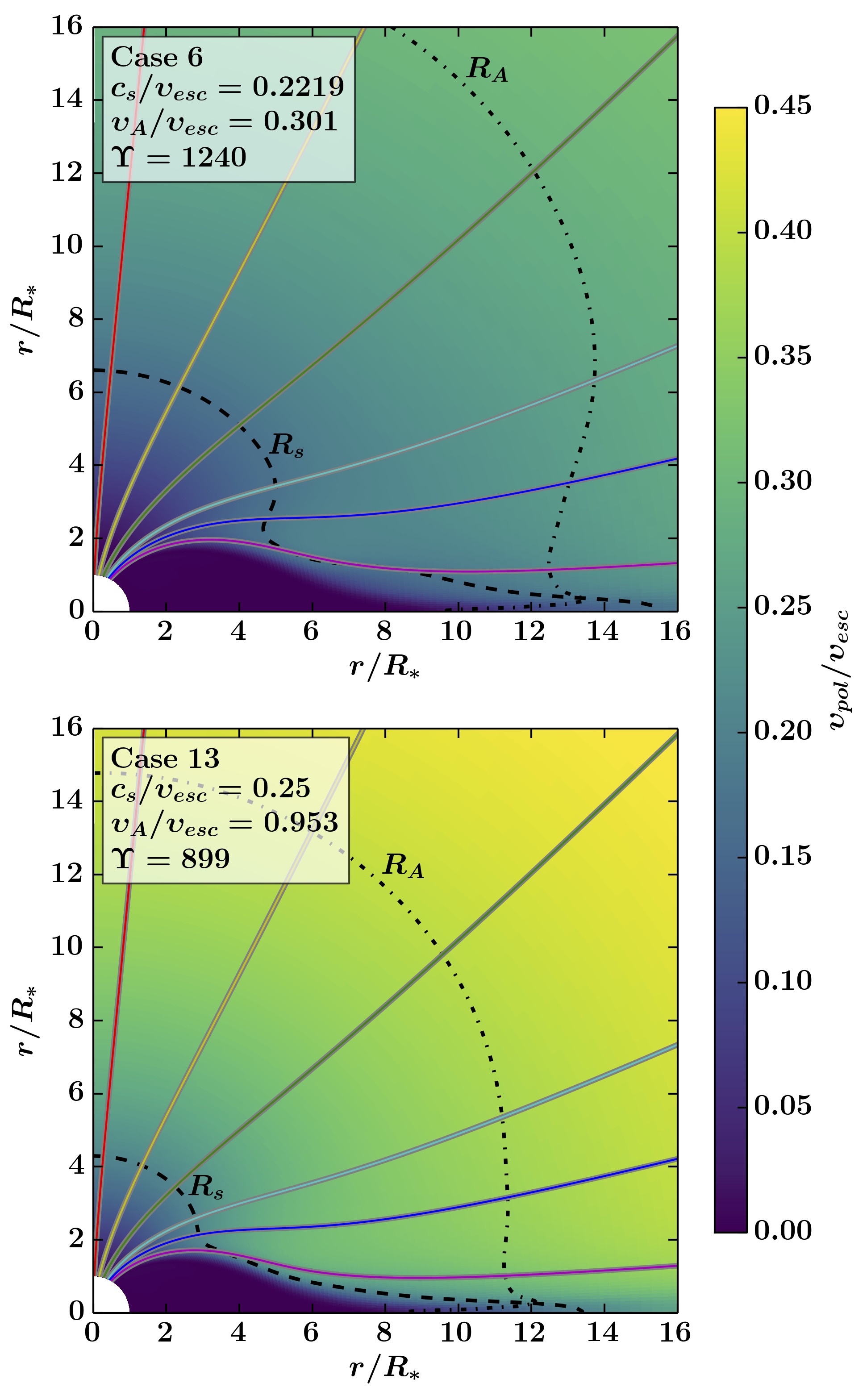}
\caption{Poloidal velocity (color scale) with magnetic field lines,
  for two steady-state wind solutions of this study that demonstrate
  the two-dimensional structure of the wind and the
  effect of the temperature on flows with similar magnetization
  (parameter $\Upsilon$) values. The dashed lines depict the sonic
  surface and the dotted lines depicts the Alfv\'{e}nic surface. Each
  field line is plotted with a different color to indicate the paths
  along the flow open streamers, plotted in figure \ref{Vpol_1d}. The
  images show only the northern stellar hemisphere and an inner
  portion of the whole computational domain.} 
\label{Vpol_2d}
\end{figure} 

For pure hydrodynamic polytropic stellar winds the two main physical
parameters that determine the wind speed and acceleration are the
temperature of the plasma and the polytropic index, $\gamma$. In this
study we focus on how different coronal temperatures affect the
driving of the outflow. The following three dimensionless 
velocities are the main input parameters of our initial setup: the
ratio of the adiabatic sound speed, defined at the stellar surface, to
the esacpe speed, $c_s/\upsilon_{esc}$, where $c_s =\sqrt{\gamma
  p_*/\rho_*}$, ("*" symbol denotes values at $R_*$), and
$\upsilon_{esc}=\sqrt{2GM_*/R_*}$; the ratio of the Alfv\'{e}n speed to
the escape speed, $\upsilon_A/\upsilon_{esc}$, where
$\upsilon_A=B_*/\sqrt{4 \pi \rho_{*}}$; the stellar spin rate, $f$,
that is the ratio of the stellar equatorial rotation velocity to the
break-up speed, where the break-up speed is
$\upsilon_{kep}=\upsilon_{esc}/\sqrt{2}$. The latter one will be held 
fixed for our study close to the solar value, $f=0.00393$. The
polytropic index $\gamma$ and the magnetic field geometry are also
parameters, but we only vary the dipolar field strengths and we fix
$\gamma = 1.05$ \citep{Washimi:1993aa,Matt:2012ab,Reville:2015ab},
which behaves like an adiabatically expanding flow that has energy
input as the wind expands, such that $p \propto
\rho^{1.05}$. 

A polytropic treatment of the outflow acceleration is suitable for our
purpose because we do not attempt to produce stellar wind solutions
that will exhibit plasma properties similar to the ones observed in
the solar wind such as speed bimodality, contrast in temperature and
density between coronal holes and helmet streamers. Regardless,
studies have shown that the polytropic approximation can capture the
large-scale structure of the solar-corona magnetic field \citep[see,
e.g.][]{Mikic:1999aa,Riley:2006aa} and produces wind solutions with
velocity profiles that agree with the observed solar wind on large
scales \citep[see, e.g.][]{Keppens:1999aa,Ofman:2004aa}.

Using the ideal-gas equation of state, the stellar coronal temperature
can be written in terms of parameter $c_s/\upsilon_{esc}$,
\begin{eqnarray}
\label{eq_coronalT}
T_* = \left( c_s \over \upsilon_{esc} \right)^2 \left( {2 G M_*
  \tilde{\mu} m_p} \over {\gamma R_* k_B} \right),
\end{eqnarray}
where $k_B$ is the Boltzmann constant, $m_p$ is the proton mass and
$\tilde{\mu}$ is the mean atomic weight (i.e. the average mass per
particle measured in units of $m_p$). For given stellar parameters,
the temperature depends on the mean atomic weight, $\tilde{\mu}$, that
is determined by the chemical composition, and the atomic physics of
the stellar atmosphere. For a solar-coronal plasma, $\tilde{\mu}=0.6$
\citep[e.g.][]{Priest:2014aa}, Table \ref{tab_coronalT} translates
$c_s/\upsilon_{esc}$ in Kelvin, for solar parameters (with $M_{\odot} =
1.99 \times 10^{33} \textrm{g}$ and $R_{\odot} = 6.96 \times 10^{10}
\textrm{cm}$), and for stars at the age of the Sun, with parameters of
$M_* = 0.7, 0.5, 0.2M_{\odot}$ and respectively $R_* = 0.65, 0.47,
0.22R_{\odot}$, taken from stellar evolution models of
\citet{Baraffe:1998aa}.

Figure \ref{HDwind_log_Vr} shows velocity profiles of polytropic
models for different coronal temperatures, represented in the plot by
the dimensionless quantity $c_s/\upsilon_{esc}$. Each curve in this
plot is the analytic solution of wind speed as a function of radial
distance from the stellar surface, and each temperature is indicated
by a different color. The plot shows that a hotter wind starts on the
stellar surface at a higher speed and also reaches a higher terminal
speed. To be more specific, for this range in $c_s/\upsilon_{esc}$,
the flow speed varies by 3.5 orders of magnitude at $R_*$, and by more
than a factor of 2 at $50R_*$. Morever, a hotter wind accelerates more
rapidly compared to a cooler wind, meaning that, at every radius, the
hotter wind exhibits a higher value of both $d \upsilon_{r}/dt$ and $d
\upsilon_{r}/dr$. Input parameter $c_s/\upsilon_{esc}$ varies between
0.2219 and 0.4, a range that was selected to produce reasonable wind
velocity profiles for the whole grid of simulations, for a given
polytropic index (i.e. $\gamma=1.05$  in our case). This range ensures
that the lowest temperature still results in a high enough flow terminal
velocity for the wind to be able to escape star's gravity field. The
upper limit for our flow temperature is determined so that it
initiates at the stellar corona  at subsonic velocities. Our wind
solution with $c_s/\upsilon_{esc}=0.4$ starts at the bottom of the
flow with an initial speed that is already $50\%$ of the sound speed,
defined at the stellar surface (see fig. \ref{HDwind_log_Vr}), and the
wind becomes supersonic at $r = 1.7R_*$. Higher temperatures will
result in outflows with unrealistically high base velocities
(i.e. almost supersonic flow at the stellar surafce). Although the
polytropic wind formalism includes simplified physics that do not
incorporate all relevant coronal processes that drive such outflows,
figure \ref{HDwind_log_Vr} shows that the range of winds we consider
in our study covers a wide range of wind acceleration profiles, which
may encompass the range of velocities encountered in real stellar
winds under various coronal conditions.

Table \ref{tab_results_a} presents the parameters varied ($2^{nd}$ and
$3^{rd}$ columns) for all the simulated wind cases in the study. The
magnetization of the wind is computed using the formula introduced in
\citet{Matt:2008aa}, 
\begin{eqnarray}
\label{eq_Upsilon}
{\Upsilon} \equiv {{B_*^{2}R_*^{2}}\over{\dot{M}_w\upsilon_{esc}}},
\end{eqnarray}  
and the quantity $\Upsilon$ can be regarded as the ratio of
the magnetic field energy to the kinetic energy of the flow, or as
representing the interplay between the Lorentz forces and the inertia
of the wind \citep{ud-Doula:2002aa}. In equation (\ref{eq_Upsilon}),
$\dot{M}_w$ is extracted dircectly from the simulations, and, for a
given surface denisty, depends on the wind-driving physics, the magnetic
field structure/configuration, and the numercial setup \citep[for
further discussion see][and subsection
\ref{subsec_OutflowRates}]{Matt:2012ab}. Therefore we choose to
present $\Upsilon$ as the second independent variable of the study,
even though $\upsilon_A/\upsilon_{esc}$ is the input parameter that
controls the magnetic field strength. All the values of $\Upsilon$ are
listed in the $4^{th}$ column in table \ref{tab_results_a}. The
parameter space that has been explored during the entire study is
visualized in figure \ref{ParamSpac}, and each simulation is one
symbol in this plot. Different symbols and their corresponding colors
repesent cases with different temperartures, and overall, we covered 3
to 4 orders of magnitude in wind magnetization for each temperature.

\subsection{Wind Velocity Profiles} \label{subsec_VelProf}

\begin{figure}
\plotone{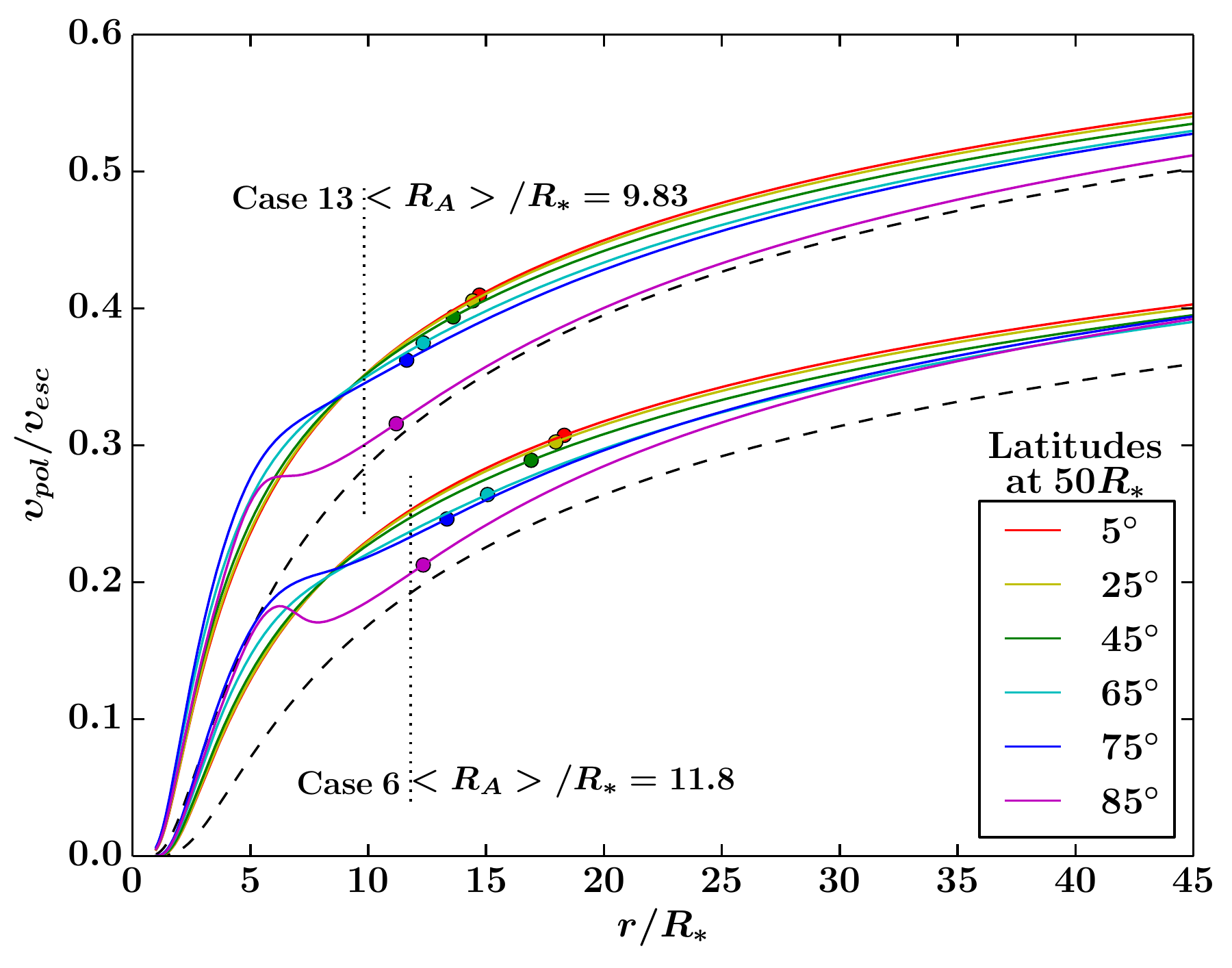}
\caption{Wind speed profiles along open field lines at different
  latitudes, as a function of radial distance, for the cases showed in
  Figure \ref{Vpol_2d}. Each line color correlates with the plotted
  field lines in Figure \ref{Vpol_2d}. For comparison, the dashed
  lines represent the velocity profiles of pure, one-dimensional
  hydrodynamic winds. The dotted lines show the torque-averaged
  Alfv\'{e}n radius or magnetic lever-arm of the magnetized outflow.}
\label{Vpol_1d}
\end{figure}

At the start of a simulation, the presence of rotation and magnetic
field modifies the initial, spherical symmetric flow, but after some
time of evolution the solution relaxes to a steay-state. In order to
highlight the influence of the gas temperature on the wind speed in
our 2.5D MHD simulations, figure \ref{Vpol_2d} shows the flow poloidal
velocity as a color scale on a subset of our domain for two
steady-sate wind solutions. Both cases shown have the same order of
magnitude in parameter $\Upsilon$. The sonic surface is notated by
$R_s$ (dashed line) and the Alfv\'{e}nic surface by $R_A$ (dot-dashed
line). Open field lines, that correspond to wind streamlines, are also
shown. A higher coronal temperature increases the velocity of the flow
(bottom panel), and as a result the sonic surface is closer to the
stellar surface. The location of the Alfv\'{e}n surface also comes
closer to the star, and this is due to a hotter and faster wind, and
also to a slightly lower magnetization of that case (i.e. case 13)
relative to top panel case (i.e. case 6).

To show how the wind velocity profile varies with latitude, figure
\ref{Vpol_1d} illustrates the poloidal speed versus radial distance,
of the plasma flowing along the streamlines of the two cases shown in
Figure \ref{Vpol_2d}. Each velocity law in figure \ref{Vpol_1d} is
individually colored, and matches the colors of the open-field
lines plotted in figure \ref{Vpol_2d}. The streamlines were chosen to
be at various latitudes at $50R_*$. The plot comprises two groups of
lines, one for each case, and the upper set correspond to the hotter
wind (i.e. case 13). Once more, it is clear that the hotter wind
accelerates more rapidly and is faster everywhere. An interesting
feature shown in Figure \ref{Vpol_1d} is that each field line
produces a unique velocity profile. This behavior should be attributed
to a different geometrical expansion of flux tubes near the pole and
close to the equator, something that originally was pointed out in
\citet{Pneuman:1971aa}. The fact that the 2D wind speed profiles are
always faster compared to their 1D hydrodynamic counterparts (black
dashed lines) occurs because of the overall, faster-than-$r^2$
divergence (i.e. superradial expansion) of the field geometry that
channels the flow
\citep[e.g.][]{Pneuman:1966aa,Kopp:1976aa,Reville:2016ab}. Since all
of our models are in the slow-magnetic-rotator regime
\citep{Belcher:1976aa}, magneto-centrifugal effects are negligible. We
verified that in the absence of rotation wind speed profiles do not
change by more than $2\%$ compared with simulated cases from rotating
stars. Furthermore, we have observed that, for everything else to be 
equal, an increase in surface field strength will produce a wind
solution that is faster everywhere, (by $\sim 10\%$), also due to
a different geometrical expansion of the flux tubes cross-section. The
circles in Figure \ref{Vpol_1d} repsresent the location of the
local Alfv\'{e}n radius, the radial distance at which the flow
along each field line reaches the local poloidal Alfv\'{e}n
speed. From stellar pole to equator the spherical Alfv\'{e}n radius
decreases because the alfv\'{e}nic surface reaches the cusp or neutral
point of the helmet streamer (closed magnetic loops), that determines
the transition region from subalfvenic to superalfvenic flows for
streamers adjacent the to the last closed field line
\citep{Pneuman:1971aa}. Finally, the black dotted vertical lines
depict the size of the  effective Alfv\'{e}n radius. The local
$R_A$ in each streamline, is always larger compared to $<R_A>$,
because the latter represents a mean value of the cylindrial
Alfv\'{e}n radius. Comapring the two cases, simulation 13 has a
smaller effective lever arm, due to both a higher coronal temperature
and a smaller $\Upsilon$ value and this yields a less efficient
braking torque on the star.

\section{Global Stellar Wind Properties} \label{sec_GlobProp}

\subsection{Mass and Angular Momentum Outflow
  Rates} \label{subsec_OutflowRates}

\begin{figure*}
\plotone{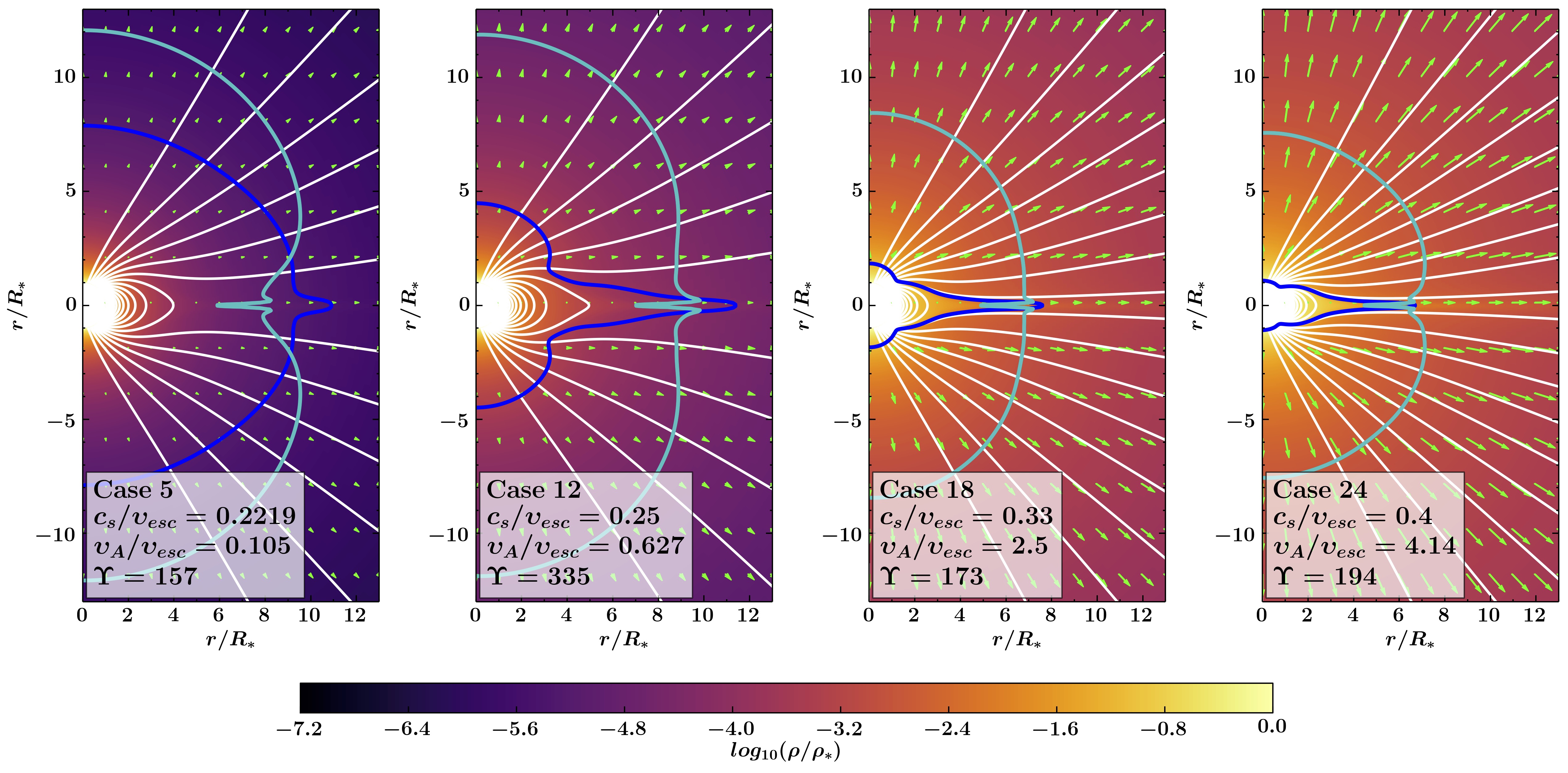}
\caption{Colormaps of logarithmic density, magnetic field
  lines, and velocity vectors, in the inner region of four simulations
  with similar magnetization, $\Upsilon$, but varying wind temperature
  (characterized by $c_s/\upsilon_{esc}$). The blue and cyan lines show
  the location of the sonic and the Alfv\'{e}nic surface, respectively. A
  higher surface plasma temperature, for about the same value of
  $\Upsilon$, results in a denser wind and the two critical surfaces
  being closer to the star.}
\label{Density_2d}
\end{figure*} 

Figure \ref{Density_2d} displays color scale plots of logarithmic
density with velocity vectors and magnetic field lines (white lines),
for 4 steady-state wind solutions of our study. Each case in figure
\ref{Density_2d} has the same order of magnitude (and about the same
value) in magnetization, but a different plasma
temperature. Qualitatively we identify that hotter winds lead to both 
a smaller sonic surface (blue line) and alfv\'{e}nic surface (cyan
line), as a consequence of being faster everywhere in the grid.

The global outflow rates of mass, $\dot{M}_w$, and angular momentum,
$\tau_{w}$, are numerically computed for each steady-state wind
solution of the study, by using  
\begin{eqnarray}
\label{eq_Mdot}
\dot{M}_w = \oint_{S} \rho \bm{\upsilon} \cdot
  d\bm{S}, 
\end{eqnarray}
\begin{eqnarray}
\label{eq_Jdot}
\tau_w = \oint_{S} \Lambda \rho \bm{\upsilon} \cdot
  d\bm{S}, 
\end{eqnarray}
where the integration occurs over any spherical surface that
encloses the star, within our computational domain, and 
\begin{eqnarray}
\label{eq_Lambda}
\Lambda = r \sin\theta \left(\upsilon_{\phi} - B_{\phi} {B_p \over
  {\rho \upsilon_p}} \right).
\end{eqnarray}
In the ideal MHD regime, $\Lambda$ gives the specific angular momentum
carried away by the wind along a streamline, and is a constant of
motion for an axisymmetric, steady-state flow. In practise, we
calculate both rates as functions of spherical radius $r$, and use the
median values obtained from all the integrated $\dot{M}_w(r)$ and
$\tau_w(r)$ over spherical shells above $10R_*$ as global
$\dot{M}_w$ and $\tau_{w}$. This method avoids numerical diffusion
effects that might cause non conservation of mass and angular momentum
flux close to the stellar boundary. We then determine the
torque-averaged Alfv\'{e}n radius, $<R_A>/R_*$, from equation
(\ref{eq_rA}), and these are listed in $5^{th}$ column of Table
\ref{tab_results_a}. 

\begin{figure}
\plotone{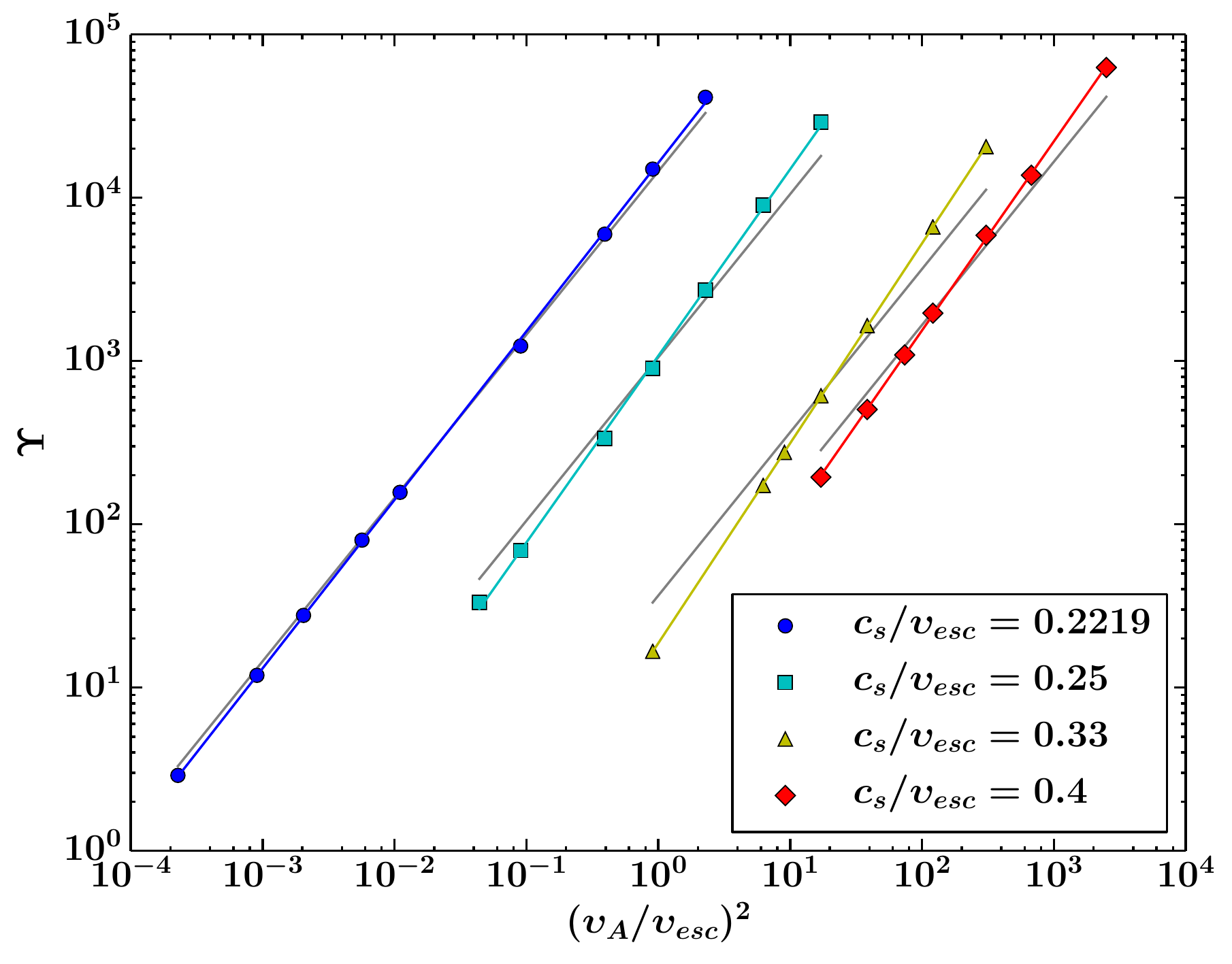}
\caption{Wind magnetization, $\Upsilon$, versus square of input
  parameter $\upsilon_A/\upsilon_{esc}$. Same colors/symbols
  correspond to a grid of simulations with the same value of
  $c_s/\upsilon_{esc}$ (as in Figure \ref{ParamSpac}). In our
  simulations, $\Upsilon \propto
  (\upsilon_A/\upsilon_{esc})^2/\dot{M}_w$, and for a given value of
  $\upsilon_A/\upsilon_{esc}$, a hotter wind has have a much higher
  mass loss rate. Grey scaling laws have a slope of unity. For a given
  coronal temperature, each scaling law has a slope steeper that
  unity, indicating that $\dot{M}_w$ decreses weakly with an
  increasing $\upsilon_A/\upsilon_{esc}$.}
\label{YvsVAVesc}
\end{figure}

Another way to illustrate the range of the parameter space is to
express $\Upsilon$ in terms of the input parameter
$\upsilon_A/\upsilon_{esc}$. By manipulating equation
(\ref{eq_Upsilon}), one can derive that $\Upsilon \propto
(\upsilon_A/\upsilon_{esc})^2/\dot{M}_w$, (i.e. $\Upsilon$ depends on
$\upsilon_A/\upsilon_{esc}$, that controls the surface magnetic field
strength, but also is inversely proportional to the stellar mass loss
rate, which is an output of the simulations). Figure \ref{YvsVAVesc}
shows that the four different temperatures of our models follow four
different scaling laws of $\Upsilon$ versus the square of
$\upsilon_A/\upsilon_{esc}$. An increase in $c_s/\upsilon_{esc}$
singnificantly affects the stellar mass-loss rates by increasing the
speed at the base of the wind. As a consequence $\Upsilon$ decreases,
and therefore we have altered the range in field strength
(i.e. variation in $\upsilon_A/\upsilon_{esc}$) for each temperature
in order to achieve about the same range in the wind magnetization for
all the temperatures. By doing this, we avoid simulations with a small
value of $\upsilon_A/\upsilon_{esc}$, and as a consequence 
a small value of $\Upsilon$, since for such cases the Alfv\'{e}n
surface is very close to the stellar surface. There is no physical
reason for not conisdering cases with wind magnetization above $10^5$,
but these simulations start to become numerically very challenging,
due to smaller numerical time-steps and large numerical errors (for
further details on the accuracy of the numerical solutions see
Appendix \ref{Appendix_Accuracy}).

The grey lines in Figure \ref{YvsVAVesc} correspond to scaling laws
with slopes of unity, and show how parameter $\Upsilon$ would depend
on $\upsilon_A/\upsilon_{esc}$, if the stellar mass loss rate was
constant for a grid of simulations with a given coronal temperature,
and thus independent of stellar surface magnetic field strength. The
fact that we find steeper power-laws, (the slopes are
respectively 1.03, 1.14, 1.22, and 1.16 for
$c_s/\upsilon_{esc}=0.2219,0.25,0.33,0.4$), indicates that the mass 
loss rates actually decrease with increased field strengths. This
feature can be physically explained by an interplay between two
competing effects. A stronger field leads to a slightly faster flow
(discussed in \S \ref{subsec_VelProf}), but also to a smaller area on
the stellar surface carrying mass flow. Figure \ref{YvsVAVesc}
indicates that the net result is a slightly decreasing $\dot{M}_w$. A
similar trend was also seen in \citet{Reville:2015ab}. Nevertheless,
we should be cautious in interpreting the scaling laws in figure
\ref{YvsVAVesc} as realistic stellar mass-loss indicators, since
polytropic wind models lack the exact physics that drive outflows from
solar- and late-type stars. Early studies in the solar wind
\citep{Leer:1980aa} showed that where the energy is added in the flow
has a big influence on the resulting solar mass loss rate. Moreover
latest theoretical models \citep{Cranmer:2011aa, Suzuki:2013aa},
suggest that a realistic treatment of coronal heating is needed for
accurate predictions on stellar mass loss rates from cool
stars. Therefore, the scaling laws between $\Upsilon$ and 
$\upsilon_A/\upsilon_{esc}$ can be interpreted as a part of the
generic phenomenology in our simulations, and should not be regarded
as trends that give accurate predictions on mass loss rates in solar-
and late-type stars. Still our formulae shall provide the exerted
magnetic torque for any given $\dot{M}_w$, extracted from observations
\citep[e.g.][]{Wood:2002aa,Wood:2014aa} or modeling
\citep[e.g.][]{Holzwarth:2007aa,Cranmer:2011aa, Suzuki:2013aa}.

\subsection{Scaling Laws Between Alfv\'{en} Radius and
  $\Upsilon$} \label{subsec_RAvsY}

\begin{figure}
\plotone{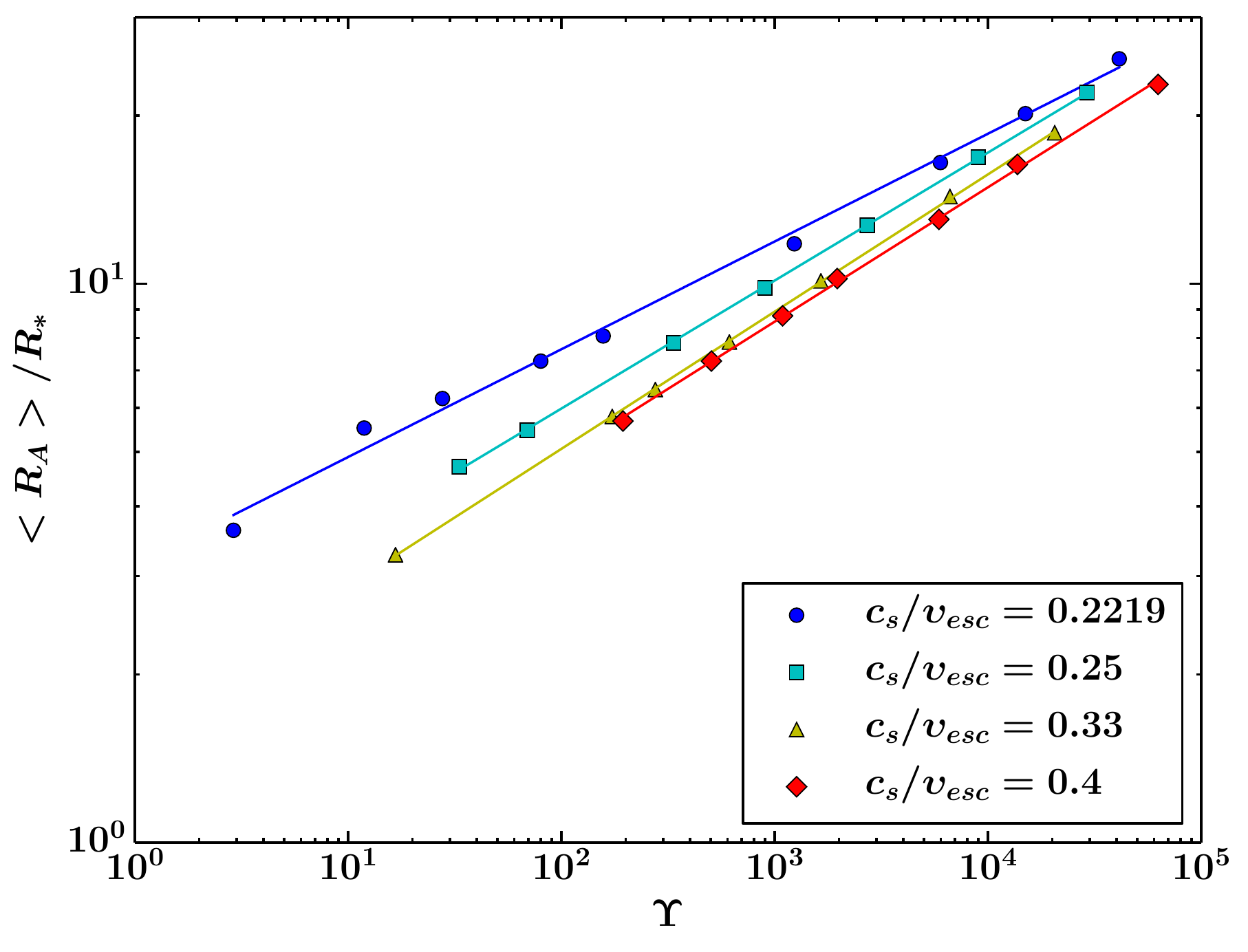}
\caption{The dependence of the effective Alfv\'{e}n radius,
  $<R_A>/R_*$, on $\Upsilon$ for all the cases of the parameter
  study. The colors/symbols have the same meaning as in Figure
  \ref{ParamSpac}. Four simple power laws of $<R_A>/R_*$ on parameter
  $\Upsilon$ are shown, and each one corresponds to a different value
  of $c_s/\upsilon_{esc}$. For a given $\Upsilon$, the magnetic lever
  arm (i.e. $<R_A>/R_*$) of the wind decreases, with an increasing
  coronal temperature, and as a consequence the torque exerted on the
  star becomes less efficient.}
\label{rAvsY}
\end{figure}

The dependence of the effective Alfv\'{en} radius, $<R_A>/R_*$ on wind
magnetization, $\Upsilon$, for all the numerical solutions of the study
is depicted in figure \ref{rAvsY}. Each point in Figure \ref{rAvsY}
corresponds to a single simulation, and the color and symbols have the
same meaning as in figure \ref{YvsVAVesc}. In order to fit the
simulation data, we use the formulation introduced in
\citet{Matt:2008aa}, that scales $<R_A>/R_*$ as a power law in 
$\Upsilon$,
\begin{eqnarray}
\label{eq_rAvsUpsilon}
{{<R_A>}\over{R_*}}=K_s \Upsilon^{m_s},
\end{eqnarray}
where $K_s$ and $m_s$ are dimensionless fitting constants, and
equation (\ref{eq_rAvsUpsilon}) determines $<R_A>/R_*$ in terms of the
magnetic field strength on the stellar surface. Four different fitting
laws are shown in Figure \ref{rAvsY}, and the values of $K_s$ and
$m_s$ for every fit are given in $2^{nd}$ and $3^{rd}$ column of Table
\ref{tab_results_b}.

Each value of $c_s/\upsilon_{esc}$ gives a simple power law of the
torque-averaged Alfv\'{e}n radius on $\Upsilon$ for various surface
magnetic field strengths. However the fit parameters are different
with each coronal temperature. The power law for $c_s/\upsilon_{esc} =
0.2219$ is shallower, (see also Table \ref{tab_results_b}), compared
with previous parameter studies \citep{Matt:2012ab, Reville:2015ab},
and can be understood as an effect due to differences in the numerical
setup between the studies (e.g. geometry of the problem, numerical
scheme, different approach on boundary conditions), indicative of
systematic errors. \citet{Reville:2015ab} demonstrated different power
laws resulted from different field geometries. It was also shown that
the compexity of the magnetic field does not significantly influence
the wind acceleration. For this study only dipolar fields are
considered, but by varying the gas temperature, we actually 
change the acceleration of the flow. As a consequence the wind speed
also changes, for simulations with different values of
$c_s/\upsilon_{esc}$, and that physically explains the four power laws
in Figure \ref{rAvsY}. In conclusion, hotter winds are faster, and
thus, the Alfv\'{e}n surface comes closer to the star, the size of the
lever arm or the effective Alfv\'{e}n radius decreases, and therefore the
magnetic braking torque that is exerted on the star becomes weaker.

\subsection{Scaling Laws Using the Amount of Open Magnetic
  Flux} \label{subsec_RAvsYopen}

In \citet{Reville:2015ab} an alternative formulation for the
torque-averaged Alfv\'{e}n radius was introduced, that scales
$<R_A>/R_*$ as a power law in a new $\Upsilon$-like parameter that
depends on the amount of open magnetic flux, \citep[see
also][]{Washimi:1993aa}. In general, the unsigned magnetic flux of the
stellar magnetic field, as a function of spherical radius $r$, can be
evaluated as, 
\begin{eqnarray}
\label{eq_PHIopen}
\Phi(r) = \oint_S \vert \bm{B} \cdot d\bm{S} \vert,
\end{eqnarray}
where the integration is performed over sperical surfaces that
enclose that star. For a given field geometry, dipole in our case,
magnetic flux intitially drops as $1/r$, but there is a regime in
which the thermal pressure and the inertia of the wind dominates over
the magnetic stresses, the field completely opens and the magnitude of 
the magnetic flux becomes constant (i.e. open magnetic flux), see for
example figure 5 in \citet{Reville:2015ab}.  

Following \citet{Reville:2015ab}, the new $\Upsilon$-like parameter,
is defined as,
\begin{eqnarray}
\label{Yopen}
{\Upsilon_{open}}\equiv{{\Phi_{open}^{2}}\over{R_{*}^{2}\dot{M}_w\upsilon_{esc}}},
\end{eqnarray}
where $\Phi_{open}$ is the open magnetic flux that is directly computed
from the numerical simulations by equation (\ref{eq_PHIopen}). We use
as $\Phi_{open}$, for a given wind solution, the median value of
$\Phi(r)$ above the corresponding $<R_A>/R_*$ of that solution, where
we have identified that magnetic flux is constant. The $6^{th}$ column
in Table \ref{tab_results_a} lists all the values of
$\Upsilon_{open}$. The $7^{th}$ column in Table \ref{tab_results_a}
contains all the values of the fractional open flux
(i.e. $\Phi_{open}$ normalized to the surface unsigned magnetic flux,
$\Phi_*$), which can be written as
$\Phi_{open}/\Phi_*=(\Upsilon_{open}/\Upsilon)^{1/2}/(4\pi)$.

\begin{figure}
\plotone{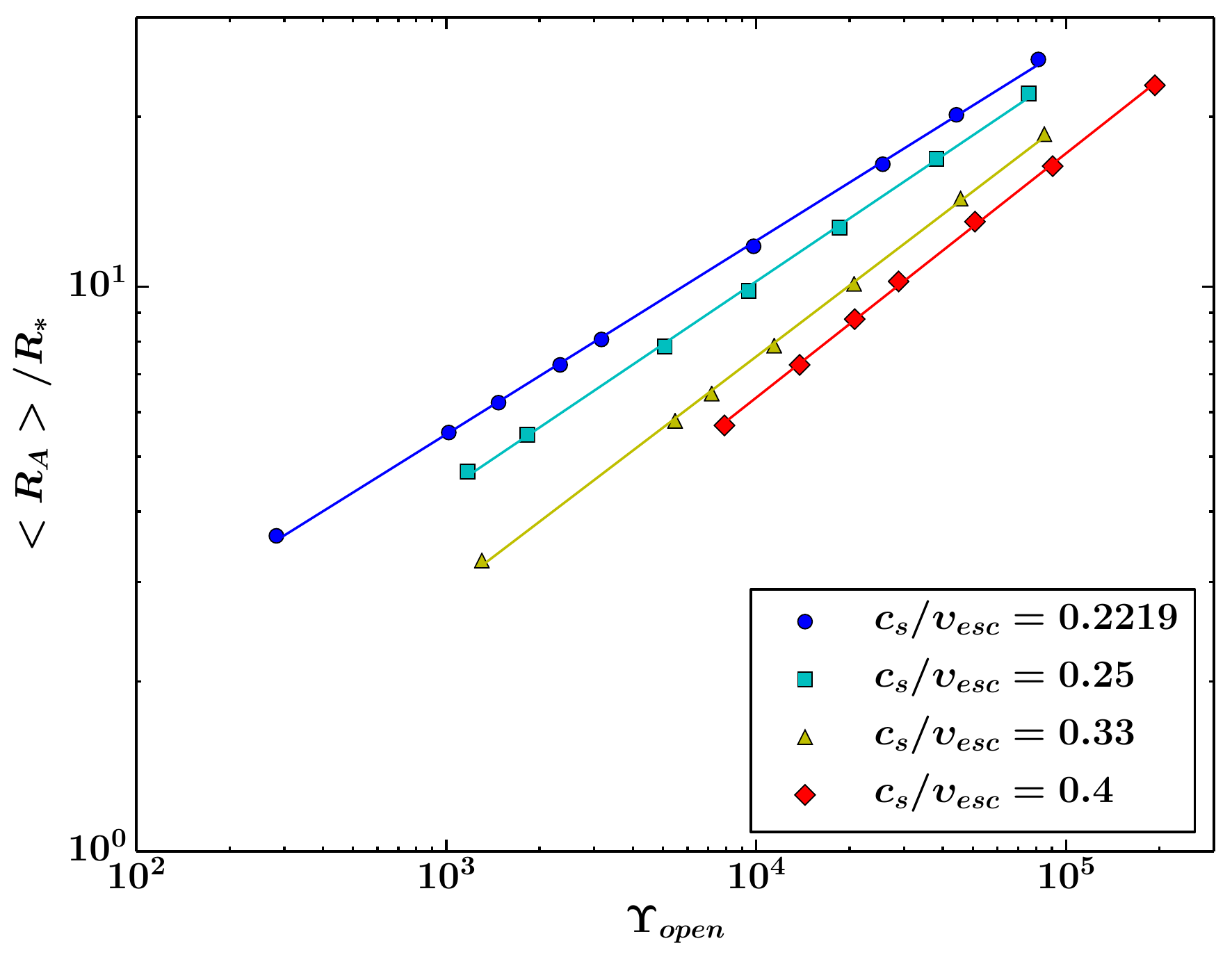}
\caption{Effective Alfv\'{e}n radius, $<R_A>/R_*$, versus the
  parameter $\Upsilon_{open}$ (eq. \ref{eq_RAvsYopen}) for all the
  simulations of the study. Colors/symbols are the same as in figure
  \ref{ParamSpac}. Four different fitting laws are shown, one for each
  set of wind solutions with a given value of $c_s/\upsilon_{esc}$. An
  increase in the temperature of the flow, for winds with the same value of
  $\Upsilon_{open}$, results in an increase of the size of $<R_A>/R_*$
  and the efficieny of the braking torque.}
\label{rAvsYopen}
\end{figure}

The value of $<R_A>/R_*$ versus parameter $\Upsilon_{open}$, for the
entire study, is presented in figure \ref{rAvsYopen}. Similarly to
equation (\ref{eq_rAvsUpsilon}), a function 
in the form of  
\begin{eqnarray}
\label{eq_RAvsYopen}
<R_A>/R_* = K_o \Upsilon_{open}^{m_o}
\end{eqnarray}
fits tha data, and again $K_o$ and $m_o$ represent dimensionless
fitting constants and $<R_A>/R_*$ is determined here in terms of the
open magentic flux. Four power laws are shown in Figure
\ref{rAvsYopen}, and the $5^{th}$ and $6^{th}$ column in Table 
\ref{tab_results_b} lists the values of the fitting constants for each
scaling law. The figure demonstrates, how effective Alfv\'{e}n radius
scales as a simple braking law with parameter $\Upsilon_{open}$, for
every value of $c_s/\upsilon_{esc}$. Furthermore, Figure
\ref{rAvsYopen} reveals one of the key result in this parameter 
study. We show that the temperature of the flow, which affects
the wind velocity and acceleration profile, is an important parameter
in the magnetic-braking models. \citet{Reville:2015ab} showed that all
the wind solutions in their study followed one unique power law,
demonstrating that the $<R_A>/R_*$-versus-$\Upsilon_{open}$ scaling
was independent of the field geometry, but they assumed a fixed
stellar coronal temperature. The fact that our power law, for
$c_s/\upsilon_{esc}=0.2219$, is steeper (see also Table
\ref{tab_results_b}), compared to the single braking law found in
\citet{Reville:2015ab}, might be explained as an effect due to
different choices in the numerical setups of the two studies, as 
discussed in the previous subsection. An influence on the braking
laws, due to a different coronal temperature has also been observed in
\citet{Reville:2016aa}. In conclusion, the temperature of the flow
affects the size of the magnetic lever-arm (i.e. $<R_A>/R_*$), and the 
efficiency of magnetic braking.

\begin{deluxetable*}{ccccccccc}
\tablewidth{0pt}
\tablecaption{Fitting Constants \tablenotemark{a} of the Parameter
  Study \label{tab_results_b}}
\tablehead{
\colhead{$c_s/\upsilon_{esc}$} &
\colhead{$K_s$} &
\colhead{$m_s$} &
\colhead{$1/(4+q)$} &
\colhead{$K_o$} &
\colhead{$m_o$} &
\colhead{$1/(2+q)$} &
\colhead{$K_q$} &
\colhead{$q$} 
}

\startdata
0.2219 & 3.1 $\pm$ 0.1 & 0.193 $\pm$ 0.005 & 0.202 $\pm$ 0.004 & 0.51
$\pm$ 0.01 & 0.343 $\pm$ 0.003 & 0.34 $\pm$ 0.01 & 0.023 $\pm$ 0.005 &
0.94 $\pm$ 0.09 \\ 
0.25 & 2.08 $\pm$ 0.02 & 0.229 $\pm$ 0.001 & 0.218 $\pm$ 0.002 & 0.34
$\pm$ 0.01 & 0.370 $\pm$ 0.004 & 0.386 $\pm$ 0.006 & 0.088 $\pm$ 0.009
& 0.59 $\pm$ 0.04 \\ 
0.33 & 1.64 $\pm$ 0.01 & 0.246 $\pm$ 0.001 & 0.230 $\pm$ 0.002 & 0.160
$\pm$ 0.007 & 0.418 $\pm$ 0.004 & 0.426 $\pm$ 0.006 & 0.32 $\pm$ 0.02
& 0.35 $\pm$ 0.03 \\ 
0.4 & 1.63 $\pm$ 0.04 & 0.240 $\pm$ 0.003 & 0.2378 $\pm$ 0.0005 &
0.118 $\pm$ 0.006 & 0.433 $\pm$ 0.005 & 0.454 $\pm$ 0.002 & 0.64 $\pm$
0.01 & 0.205 $\pm$ 0.009 \\ 
%\hline
\hline \\
0.2219 \tablenotemark{b} & 2.49 & 0.2177 & - & - & - & -
& - & - \\
0.2219 \tablenotemark{c} & 2.0 $\pm$
0.1 & 0.235 $\pm$ 0.007 & 0.21 & 0.65 $\pm$ 0.05 & 0.31 $\pm$ 0.02 &
0.37 & - & 0.7\\
\enddata
\tablenotetext{a}{For equations (\ref{eq_rAvsUpsilon}),
  (\ref{eq_RAvsYopen}), and (\ref{eq_VrA}) in Figures \ref{rAvsY},
  \ref{rAvsYopen}, and \ref{VrAvsRA}, respectively.}
\tablenotetext{b}{\citet{Matt:2012ab}}
\tablenotetext{c}{\citet{Reville:2015ab,Reville:2016aa}}

\end{deluxetable*}

\section{Magnetic braking laws for known wind acceleration
  profile} \label{sec_BrakingLaws}

\subsection{Semi-analytic Model for Alfv\'{e}n Radius versus
  $\Upsilon_{open}$} \label{subsec_1D_RAvsYopen}

\begin{figure}
\plotone{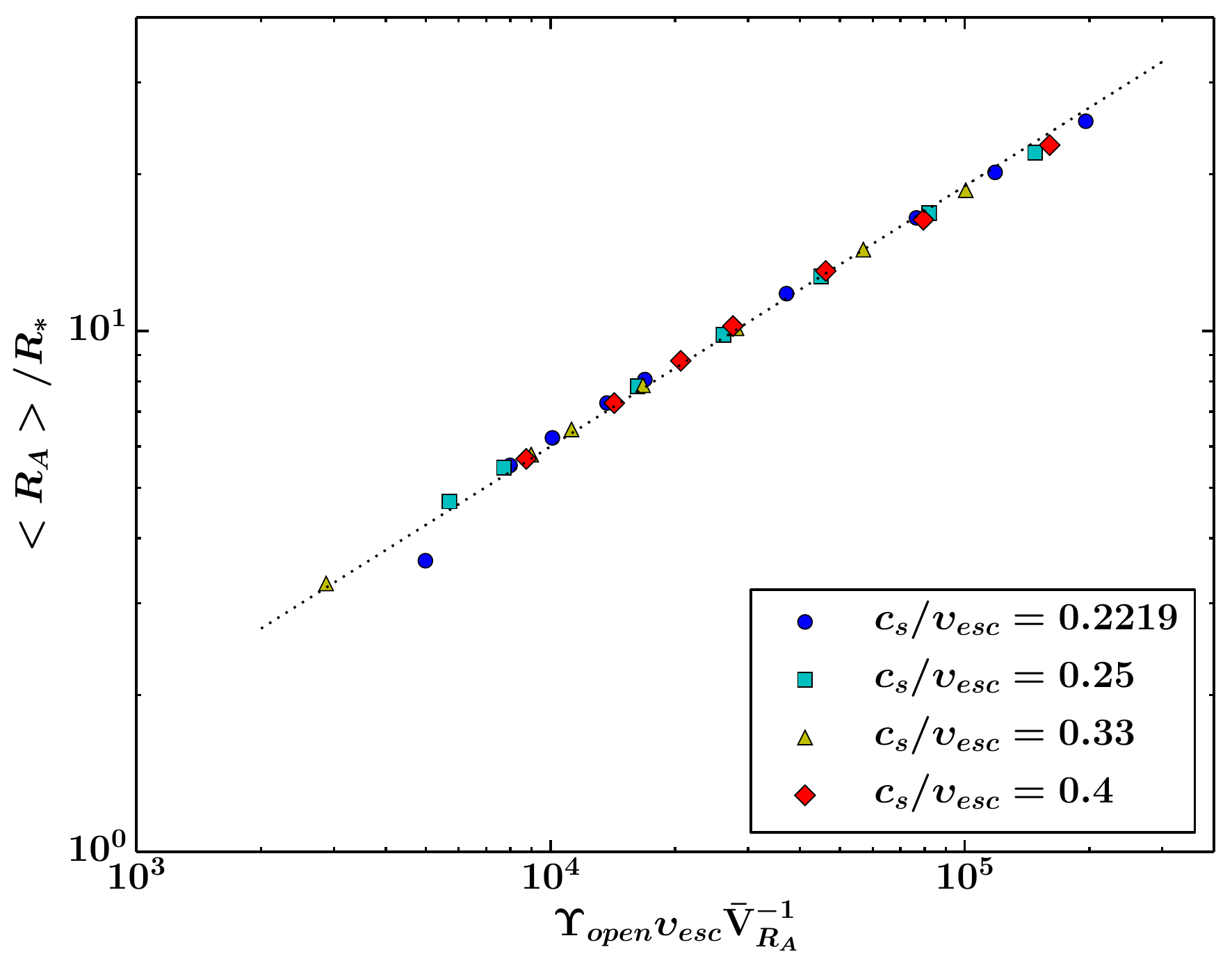}
\caption{Effective Alfv\'{e}n radius, $<R_A>/R_*$, versus the quantity
  $\Upsilon_{open} \upsilon_{esc} \bar{\mathrm{V}}_{R_A}^{-1}$ for all the
  simulation data. Colors/symbols have the same meaning as in Figure
  \ref{ParamSpac}. All the data points collapse in a single braking 
  law, compared to Figure \ref{rAvsYopen}. The slope (or power-law
  index) of the dotted line is fixed to $1/2$, and fits the data
  according to equation (\ref{eq_rAvsYopenmod_a}).}
\label{rAvsYopenmod}
\end{figure}

We showed above that the flow temperature and the resulting wind
acceleration can influence the effieciency of the braking toque. For
this section, our objective is to provide a more generic braking law
that will take this effect into account.

In order to mathematically express the dependence of the braking laws
on the acceleration profile of the flow, we will employ similar 
one-dimensional analysis that was used in earlier works
\citep[e.g][]{Kawaler:1988aa,Tout:1992aa,Matt:2008aa,Reville:2015ab}. For
a one-dimensional, MHD flow, along a magnetic flux tube, the wind
velocity at the Alfv\'{e}n radius, by defintion, is equal to the
local Afv\'{e}nic speed. This is
\begin{eqnarray}
\label{eq_VA}
\upsilon^2(R_A) = \upsilon_A^2 = {B_A^2 \over {4\pi\rho_A}},
\end{eqnarray}
where $B_A$ and $\rho_A$ are the local magnetic field and density
respectively, at the Alfv\'{e}n surface. In order to evaluate  $B_A$
at $R_A$, one must specify how the magnetic field stength depends on
radius. Hence, for this work, we adopt a prescription similar to
\citet[][see also \citet{Mestel:1999aa}]{Mestel:1987aa}, in which the
magnetic field is approximated as having two regions. The inner region
exists from the stellar surface out to the "open-field'' radius,
$R_o$, in which the field is a single power law in radius,
\begin{eqnarray}
\label{eq_Bfield_Rstar}
B(r \leq R_o) = B_* \left( {R_* \over r} \right)^{l+2},
\end{eqnarray}
with $l=1$ for a dipole. The outer region lies above $R_o$ in which
the field decreases as a monopole, (i.e. $l=0$), 
\begin{eqnarray}
\label{eq_Bfield_Ropen}
B(r \geq R_o) = B_o \left( {R_o \over r} \right)^2,
\end{eqnarray}
where $B_o$ denotes $B(R_o)$, given by equation
(\ref{eq_Bfield_Rstar}). We also assume that the flow is the same
along every field line (i.e. all values are only a function of radius
and not latitude) and in a steady-state. 

This treatment for the stellar field magnetic is a simple
approximation for the real magnetic field configurations in a wind,
where near the star, the field closely resembles the potential field,
and further out, it is stretched to a nearly radial configuration by
the flow (see for example fig. \ref{Density_2d}). For a detailed
comparison of the magnetic field in a wind simulation with a potential
and radial field, see \citet{Reville:2015aa}. 

In all our simulations the Alfv\'{e}n surface is located at the
open-field region, and therefore, we assume that the condition $R_A >
R_o$ holds for all our cases as if they were 1D flows. Then, by
combining equations (\ref{eq_Bfield_Rstar}) and
(\ref{eq_Bfield_Ropen}), the magnetic field strength at $R_A$ can now
be written 
\begin{eqnarray}
\label{eq_Bfield_RA}
B_A = B_o \left( {R_o \over R_A} \right)^2 = B_* \left( {R_* \over
  R_o} \right)^{l+2} \left( {R_o \over R_A} \right)^2. 
\end{eqnarray}
Since magnetic flux is conserved, it can be written at the Alfv\'{e}n
radius as
\begin{eqnarray}
\label{eq_MagFlux_Ro}
\Phi_A= 4 \pi R_A^2 B_A = 4 \pi R_o^2 B_o = \Phi_{open},
\end{eqnarray}
which equals the total open flux in the wind. By combining equations
(\ref{Yopen}), (\ref{eq_VA}), and (\ref{eq_MagFlux_Ro}), we get
\begin{eqnarray}
\label{eq_rAvsYopen_b}
\left(R_A \over R_* \right)^2 = {1 \over (4 \pi)^2} \Upsilon_{open}
  {\upsilon_{esc} \over \upsilon(R_A)}, 
\end{eqnarray}
where we have used $\dot{M}_w = 4 \pi \rho_A R_A^2 \upsilon(R_A)$, for
a spherical symmetric flow in the open-field region.

Since our wind solutions are multi-dimensional, we can associate the
terms $R_A/R_*$ and $\upsilon(R_A)$ in equation (\ref{eq_rAvsYopen_b})
with the torque-averaged Alfv\'{e}n radius, $<R_A>/R_*$ and
$\bar{\mathrm{V}}_{R_A}$, where $\bar{\mathrm{V}}_{R_A}$ represents
the average wind speed at the Alfv\'{e}n surface. We define
\begin{eqnarray}
\label{eq_meanVRA}
\bar{\mathrm{V}}_{R_A} \equiv {\sum \limits_{i}^{N} \upsilon[(R_A)_i,
  \theta_i], \over N}, 
\end{eqnarray}
where the sum is over each discretized grid point $i$ along the
Alfv\'{e}n surface. $\bar{\mathrm{V}}_{R_A}$ is computed individually for each
case in the study, and the values are listed in the $8^{th}$ column in
Table \ref{tab_results_a}.

Following equation (\ref{eq_rAvsYopen_b}), we plot $<R_A>/R_*$ versus
the new quantity, $\Upsilon_{open} \upsilon_{esc}
\bar{\mathrm{V}}_{R_A}^{-1}$, as depicted in Figure \ref{rAvsYopenmod}, and
fit the data to the function 
\begin{eqnarray}
\label{eq_rAvsYopenmod_a}
{<R_A> \over R_*} = K_{c} \left(\Upsilon_{open}
  {\upsilon_{esc} \over \bar{\mathrm{V}}_{R_A}} \right)^{1/2},
\end{eqnarray}
where again $K_{c}$ is introduced as a dimesionless fitting constant
and its value should only deviate from $1/(4\pi)$ due to 2D effects,
neglected in equation (\ref{eq_rAvsYopen_b}). The best-fit value for
$K_c$ gives 
\begin{eqnarray}
\label{K_c}
K_c = 0.7540 \left(1 \over 4 \pi \right) \pm 0.0004.
\end{eqnarray}
By including in our torque formalism, the dimensionless term
$\upsilon_{esc} / \bar{\mathrm{V}}_{R_A}$, that contains all the information
regarding the velocity and acceleration profile of the outflow, all
the data points in figure \ref{rAvsYopenmod} collapse in one single 
and precise power-law. Hence, equation (\ref{eq_rAvsYopenmod_a})
predicts the effective Alfv\'{e}n radius of any wind, as long as
$\bar{\mathrm{V}}_{R_A}$ and $\Upsilon_{open}$ are known.

\subsection{Power-law Approximation for Wind Velocity at the
  Alfv\'{e}n Radius, $\bar{\mathrm{V}}_{R_A}$} \label{subsec_VRA}

\begin{figure}
\plotone{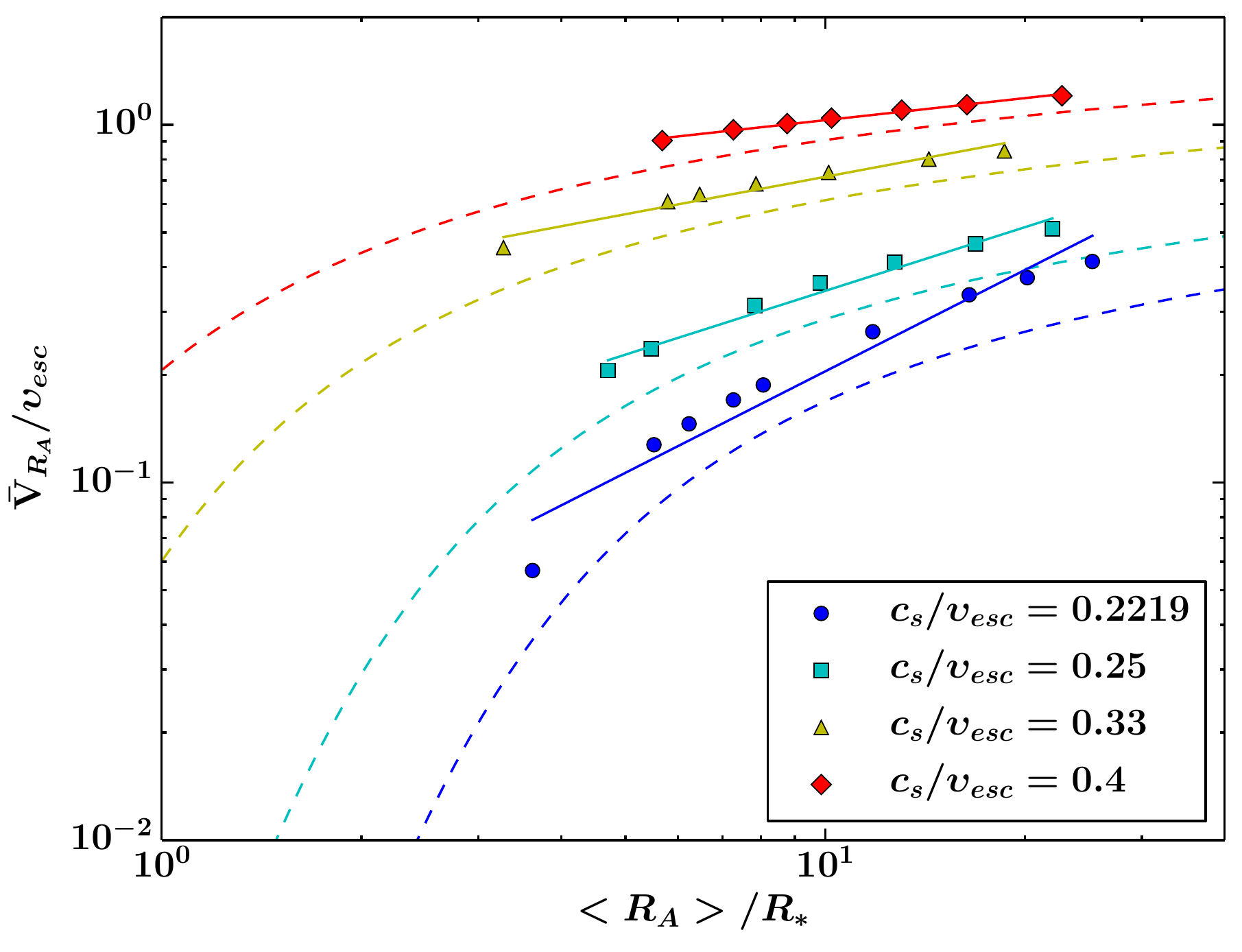}
\caption{Average flow speed at the Alfv\'{e}n surface,
  $\bar{\mathrm{V}}_{R_A}$, versus $<R_A>/R_*$ for all the simulated
  cases of the study. Colors/symbols are the same as in Figure
  \ref{ParamSpac}. Each point in this plot represents the average wind
  speed at the Alfv\'{e}n radius of a single wind solution
  (eq. \ref{eq_meanVRA}). The solid lines represent the equation
  (\ref{eq_VrA}) with fit parameters listed in Table
  \ref{tab_results_b}. For comparison, the dashed lines show the
  normalized radial velocity, $\upsilon_{r}/\upsilon_{esc}$, as a
  funcion of $r/R_*$, of the 1D, hydrodynamic, winds illustrated in
  Figure \ref{HDwind_log_Vr}.} 
\label{VrAvsRA}
\end{figure}

Equation (\ref{eq_rAvsYopenmod_a}) can naturally explain the simple
power laws in figure \ref{rAvsYopen}, if wind speed,
$\bar{\mathrm{V}}_{R_A}$, is also a power-law in $<R_A>/R_*$ but with
a scaling that varies for each temperature. To verify this, we plot
$\bar{\mathrm{V}}_{R_A}$ versus the torque-averaged Alfv\'{e}n radius,
$<R_A>/R_*$, for all the simulations in figure \ref{VrAvsRA}. For
comparison, the velocity profiles of the polytropic, Parker wind
models, shown in Figure \ref{HDwind_log_Vr}, are also plotted. We fit
a power-law  function to the data, given by 
\begin{eqnarray}
\label{eq_VrA}
{\bar{\mathrm{V}}_{R_A} \over \upsilon_{esc}} = K_q \left(<R_A> \over R_*
  \right)^q 
\end{eqnarray}
where $K_q$ and $q$ are both dimensionless fitting constants, related
to the acceleration profile of the wind. Each temperarture gives us a
seperate pair of $K_q$ and $q$, tabulated in the $8^{th}$ and $9^{th}$
column of Table \ref{tab_results_b}, respectively. The value of $q$,
found in \citet{Reville:2015ab}, is also given in Table
\ref{tab_results_b}.

It is clear that equation (\ref{eq_VrA}) is valid as a first order
approximation, despite the fact that the simulated winds do not follow
a perfect power law (solid lines in fig. \ref{VrAvsRA}) and the
behavior of $\bar{\mathrm{V}}_{R_A}$, as a function of $<R_A>/R_*$, exhibit a
similar shape to 1D, hydrodynamic winds of the same value of
$c_s/\upsilon_{esc}$ (dashed lines in fig. \ref{VrAvsRA}). Perhaps,
for even more precise stellar-torque formulae, a different velocity
law could be applied (e.g. modified beta-law, see for example
\citet{Lamers:1999aa}). Nonetheless, over a small
range of radii, these trends can be approximated by a power law, and
that approximate fit, explains the power-law behavior in figure 
\ref{rAvsYopen}. In addition, working with equation (\ref{eq_VrA}),
one can analytically solve equation (\ref{eq_rAvsYopenmod_a}) for
$<R_A>/R_*$, (see below).

Another interesting trend in figure \ref{VrAvsRA} is that the plotted
data points are noticeably above the hydrodynamic wind velocity
profiles. This can be understood as an effect due to both, the
differences in the dynamics of the two flows (i.e. MHD versus HD flow)
as discussed in section \ref{subsec_VelProf}, and the specific way the
averaging and the scaling was done in equation (\ref{eq_VrA}). Figure
\ref{VrAvsRA} also indicates why the braking laws in figures
\ref{rAvsY} and \ref{rAvsYopen} start to converge, for higher coronal 
temperatures (e.g. the yellow and red lines with
$c_s/\upsilon_{esc}=0.33,0.4$). Hotter flows enter the regime where 
the wind speed starts to saturate to wind terminal speed (i.e. speed
at infinity), in a shorter radial distance compared to cooler
winds. Hence, outflows that approach an almost constant speed, suggest
a $q$ that asymptotes to zero. Lastly, we found two empirical
functions, which predict fitting constants $K_q$ and $q$ over any
continuous range of values of $c_s/\upsilon_{esc}$. These functions are,
\begin{eqnarray}
\label{eq_Kq_full}
K_q = 1.36 [5.87  (c_s/\upsilon_{esc})^2 - 1.18
  (c_s/\upsilon_{esc})], 
\end{eqnarray}
\begin{eqnarray}
\label{eq_q_full}
q = 0.932 [0.000979 (c_s/\upsilon_{esc})^{-4.51} \nonumber \\+\ 0.553
  (c_s/\upsilon_{esc})].
\end{eqnarray}
The method and the derivation of equations (\ref{eq_Kq_full}) and
(\ref{eq_q_full}) exist in Appendix \ref{Appendix_Kq_q_vs_CsVesc}.

By combining equations (\ref{eq_rAvsYopenmod_a}) and (\ref{eq_VrA}),
we obtain
\begin{eqnarray}
\label{eq_rAvsYopenmod_b}
{<R_A> \over R_*} = \left({K_c^2 \over  K_q} \Upsilon_{open} \right)
  ^{1/(2+q)}. 
\end{eqnarray}
An interesting characteristic of equation (\ref{eq_rAvsYopenmod_b}) is
that it explains the fitting constants of equation
(\ref{eq_RAvsYopen}) in terms of other fitting constants, and consists
of an analytic expression for the effective Alfv\'{e}n radius. This
formalism is independent of the temperature of the flow (but requires
a known wind acceleration profile), the geometry of the magnetic
field, and predicts the torque exerted on the star for any value of
$\Upsilon_{open}$, for a given rotation rate (in the slow-rotator
regime) and polytropic index ($\gamma=1.05$ in this study). Comparing
equations (\ref{eq_RAvsYopen}) and (\ref{eq_rAvsYopenmod_b}), we
identify that $K_o \sim (K_c^2 K_q^{-1})^{1/(2+q)}$ and $m_o \sim
1/(2+q)$. The predicted values of $m_o$ for each temperature, are
listed in the $7^{th}$ column in Table \ref{tab_results_b}. Clearly
$m_o$ and $K_o$ stongly depend on the accelaration profile of the
wind, here parametrized with $K_q$ and $q$.

\subsection{Semi-analytic Model for Alfv\'{e}n Radius versus
  $\Upsilon$} \label{subsec_1D_RAvsY}

\begin{figure}
\plotone{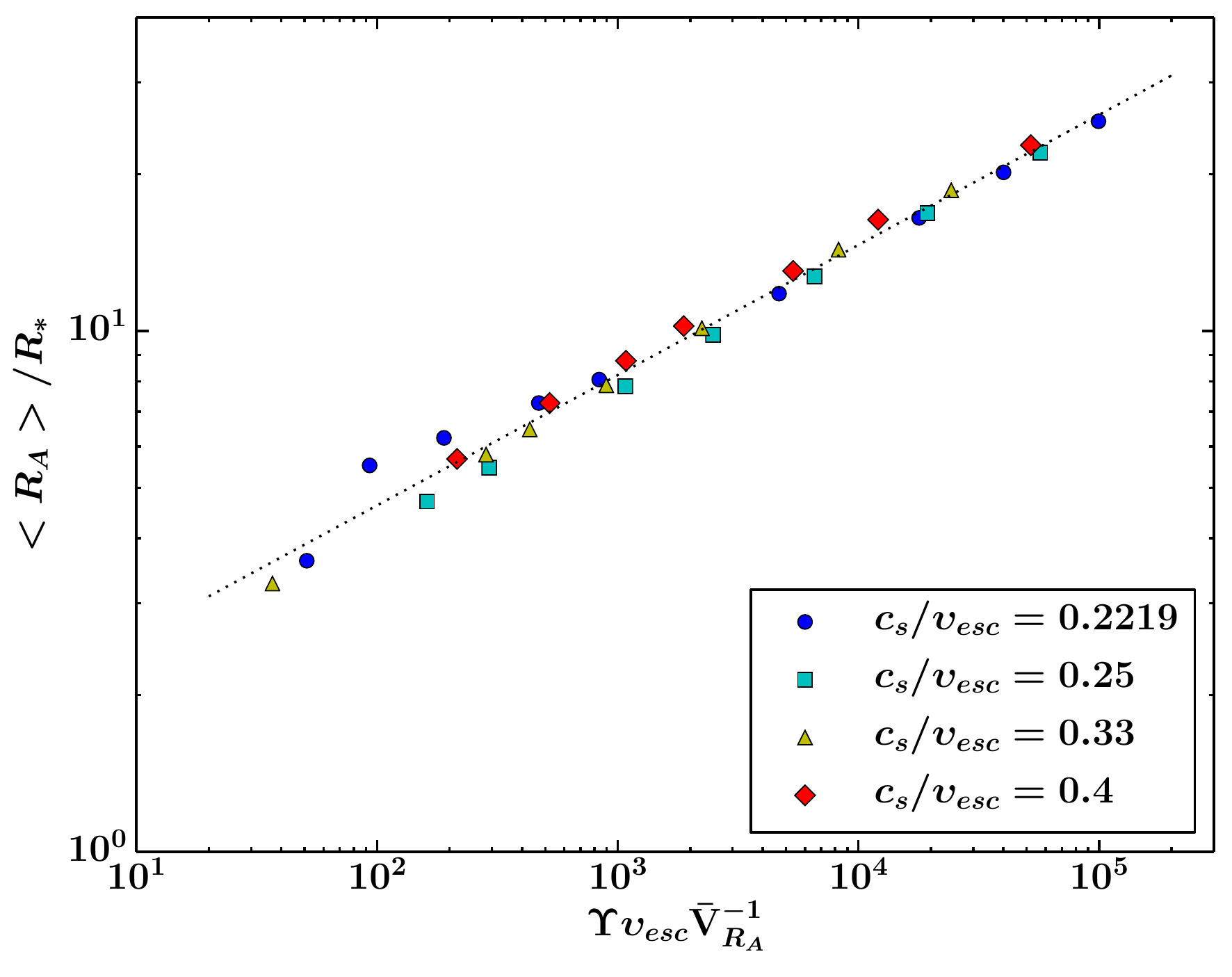}
\caption{$<R_A>/R_*$ versus the quantity $\Upsilon \upsilon_{esc}
  \bar{\mathrm{V}}_{R_A}^{-1}$ for all the simulations. Colors/symbols
  have the same meaning as in Figure \ref{ParamSpac}. All the data
  points are fitted by a single coefficient $K_l$, and the fitting
  line (dotted line) has a slope (or power-law index) of $1/4$,
  according to equation (\ref{eq_rAvsYmod_a}). The small spread of the
  data points, observed in this braking law is primarily due to
  variations in the ratio of the Alfv\'{e}n radius to the open-field
  radius (see also eq. \ref{eq_rAvsUpsilon_b} and Figure
  \ref{Ro_vs_Ra}).}
\label{rAvsYmod}
\end{figure}

The formalism given by equation (\ref{eq_rAvsYopenmod_a}) provides an
excellent fit, in terms of predicting the torque-averaged Alfv\'{e}n
radius from parameter $\Upsilon_{open}$, for a given wind
acceleration. However, in real wind cases, the amount of open magnetic
flux is a quantity that is not observable, and can only be predicted
\citep[e.g.][]{Vidotto:2014aa,Reville:2015aa,See:2017aa}. Therefore, 
in this section, we aim at extracting trends for the braking torque
based on $\Upsilon$, that depends on the surface magnetic field
strength (or surface magnetic flux).

Such trends can be obtained analytically, by combining equations
(\ref{eq_Bfield_RA}), (\ref{eq_MagFlux_Ro}), (\ref{eq_rAvsYopen_b}),
and also by using the definition for $\Upsilon$,
(eq. \ref{eq_Upsilon}), which yields
\begin{eqnarray}
\label{eq_rAvsUpsilon_b}
\left( R_A \over R_* \right)^{2l+2}  \left( R_o \over R_A \right)^{2l}
  = \Upsilon {\upsilon_{esc} \over \upsilon(R_A)}.
\end{eqnarray}

Figure \ref{rAvsYmod} shows the effective Alfv\'{e}n radius versus the
$\Upsilon$-based quantity, $\Upsilon \upsilon_{esc}
\bar{\mathrm{V}}_{R_A}^{-1} $, as it is suggested by equation
(\ref{eq_rAvsUpsilon_b}). Once more, all the details regarding the
acceleration of the flow have been contained in the dimensionless 
term, $\upsilon_{esc} / \bar{\mathrm{V}}_{R_A}$, and as a result all the
simulations lie close to a single power law. By solving equation
(\ref{eq_rAvsUpsilon_b}) for $R_A/R_*$, the power on this braking law
depends only the geometry of the field (or $l$). Hence, this should
apply to more complex field geometries as well, but for our case,
with a dipole field ($l=1$), the slope, of the single line formed by
the data points in Figure \ref{rAvsYmod}, is equal to $1/4$. Following
this simplified analysis, we fit the data in Figure \ref{rAvsYmod} with
\begin{eqnarray}
\label{eq_rAvsYmod_a}
{<R_A> \over R_*} = K_{l} \left( \Upsilon {\upsilon_{esc} \over
  \bar{\mathrm{V}}_{R_A}} \right)^{1/4},
\end{eqnarray}
and $K_l$ is introduced as the only fitting constant. The best-fit
value of $K_l$ is
\begin{eqnarray}
\label{K_l}
K_l = 1.46 \pm 0.02.
\end{eqnarray}

Fitting constant $K_l$ includes any factors, due to the
multidimensionality of our simulations, and most important, the
comparison between equations (\ref{eq_rAvsUpsilon_b}) and
(\ref{eq_rAvsYmod_a}) suggests that $K_l$ also includes the
dimensionless ratio of the Alfv\'{e}n radius to the open-field radius
of the wind, $R_A/R_o$. Furthermore, the fact that all the data points
do not precisely lie along the single power law in Figure
\ref{rAvsYmod}, implies that the term $R_A/R_o$ is not constant for
all the simulations and exhibits a dependence on the flow
temperature.

\begin{figure}
\plotone{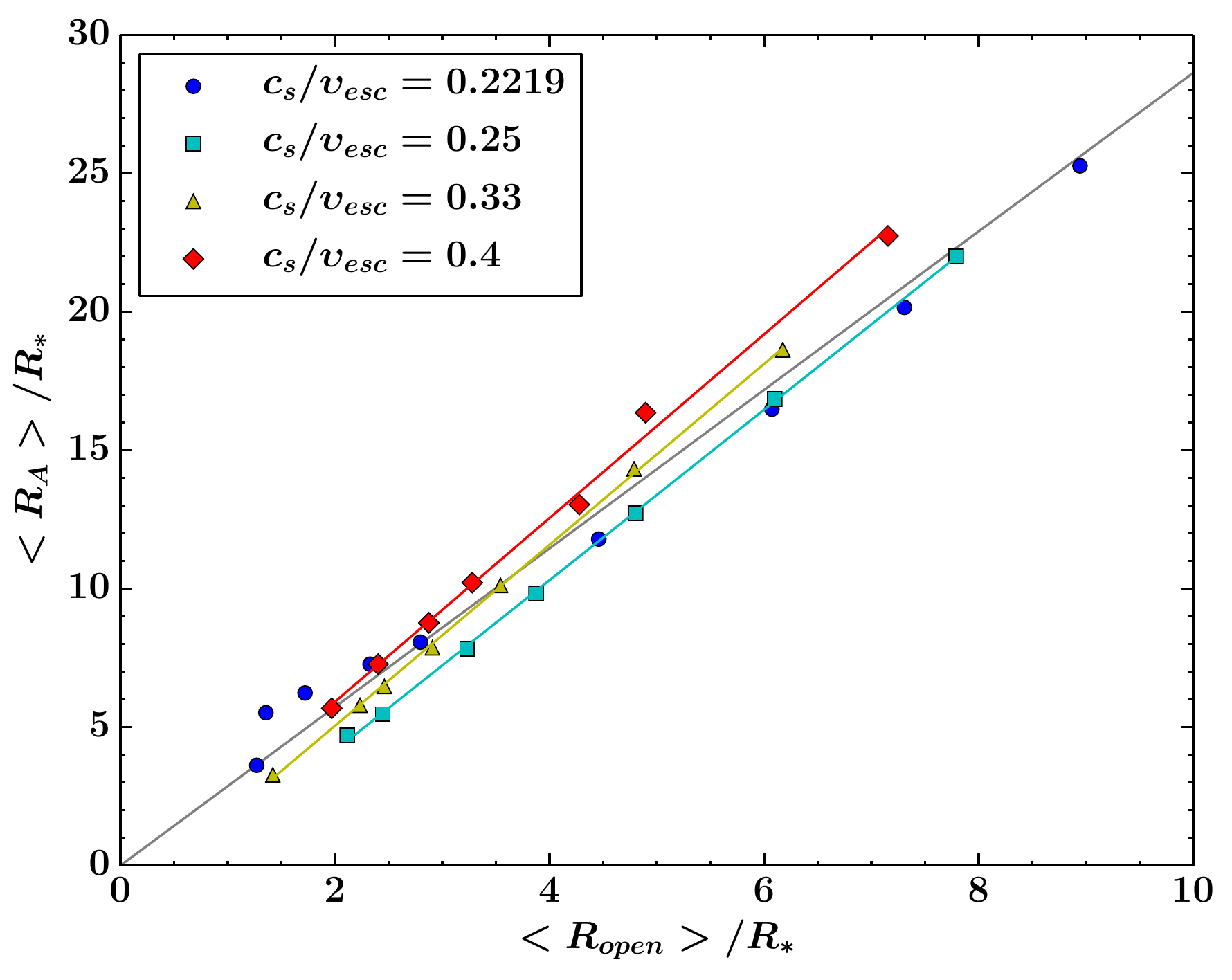}
\caption{Torque-averaged Alfv\'{e}n radius, $<R_A>/R_*$, versus the
  normalized open-field radius, $<R_o>/R_*$. Color/symbols are the
  same as in figure \ref{ParamSpac}. The grey line shows a linear
  function that represents all the data, and gives
  $<R_A>/<R_o>=2.86$. Cyan, yellow, and red solid lines depict
  linear functions as well, as an example, to show how $<R_A>/<R_o>$
  systematically varies for each temperature.}
\label{Ro_vs_Ra}
\end{figure}

A coherent way to estimate the open-field radius (i.e. the radial
distance in which the wind's thermal and ram pressure overpower the
magnetic field pressure, and as a result the unsigned magnetic flux
becomes constant as a function of radial distance) for all our
simulations, is to define $<R_o>/R_*$ as
\begin{eqnarray}
\label{eq_Ropen}
\left( {<R_o> \over R_*} \right)^l \equiv {\Phi_* \over \Phi_{open}}.
\end{eqnarray}
In other words, equation (\ref{eq_Ropen}) gives the radial distance in
which the function, $\Phi(r)/\Phi_* = (R_*/r)^l$, intersects the line,
$\Phi(r)/\Phi_* = \Phi_{open}/\Phi_* = \textrm{const.}$, and applies
for any given single magnetic field geometry.

In Figure \ref{Ro_vs_Ra}, we present the normalized open-field radius,
$<R_o>/R_*$, versus the torque-averaged Alfv\'{e}n radius,
$<R_A>/R_*$, and the plot shows that all the simulations have
approximately the same ratio, $<R_A>/<R_o>$. This feature explains why
equation (\ref{eq_rAvsYmod_a}) successfully represents the data. Assuming a
linear scaling between $<R_A>$ and $<R_o>$, yields $<R_A>/<R_o>
\approx 2.86$. A closer inspection reveals a range 
in $<R_A>/<R_o>$ between 2.23 and 4.07, that will produce a scatter in
figure \ref{rAvsYmod} only as the square root of this ratio, with the
most extreme deviation from the linear function (grey line) to be
$20\%$. In fact, $<R_A>/<R_o>$ systematically changes, which explains
the systematic scatter in figure \ref{rAvsYmod} as due to small
differences in $<R_A>/<R_o>$ for each temperature. The general trend
in figure \ref{Ro_vs_Ra}, is that $<R_A>/<R_o>$ increases for an
increasing flow temperature (see solid cyan, yellow, and red lines),
though that is not the case for simulations with
$c_s/\upsilon_{esc}=0.2219$, for which the data points exhibit a
peculiar behavior. Lastly, $<R_A>/<R_o>$ should exhibit a dependence
on the geometry of the field, and in particular, the expected trend is
that for an increasing complexity in the field geometry, this ratio
reduces \citep[see][]{Finley:2017aa}.

Equation (\ref{eq_rAvsYmod_a}) can be further expanded by substituting
$\bar{\mathrm{V}}_{R_A}$ with the velocity law, given by equation
(\ref{eq_VrA}), which yields
\begin{eqnarray}
\label{eq_rAvsYmod_b}
{<R_A> \over R_*} = \left( {K_{l}^4 \over K_q} \Upsilon
  \right)^{1/(4+q)}. 
\end{eqnarray}
Equation (\ref{eq_rAvsYmod_b}) explains the fitting constants of
equation (\ref{eq_rAvsUpsilon}) in terms of other fitting constants,
and represents an analytic formula of the torque-averaged Alfv\'{e}n
radius for any value of parameter $\Upsilon$, for any known wind
acceleration profile (known values of $K_q$ and $q$), for a dipolar
field geometry, for a star that is a slow rotator, and for
$\gamma=1.05$. 

Finally, equation (\ref{eq_rAvsYmod_b}) can work as a proxy in order to
extract predictions for the values of $K_s$ and $m_s$ (see
eq. \ref{eq_rAvsUpsilon}), that determine the simple power laws in
figure \ref{rAvsY}. It is expected that $K_s \sim (K_l^{4} /
K_q)^{1/(4+q)}$ and $m_s \sim 1/(4+q)$, see for example the $4^{th}$
column in Table \ref{tab_results_b}, for the predicted values of
$m_s$, for each flow temperature. Undoubtedly, the primary reason for
the differences in the four different power laws in figure
\ref{rAvsY}, is related to the acceleration of the flow, which depends
on the stellar coronal temperature.

\section{Summary and Conclusions} \label{sec_Conclusions}

Employing 2.5D, ideal MHD, axisymmetric numerical simualtions, we
provide the first systematic study on how the thermodynamic conditions
(i.e. flow temperature for the current work), in stellar coronae of
cool stars, can influence the losses of stellar angular momentum due
to magnetized winds. Our parameter space considers polytropic flows,
modified with rotation and magnetic fields, includes 30 steady-state
wind solutions (see Appendix \ref{Appendix_Sims} for color scale plots
of the complete simulation grid), and quantifies the braking torque
for 4 different coronal temperatures, over a wide range of magnetic
field strengths, for slow rotators, for dipolar fields, and for a
fixed polytropic index ($\gamma=1.05$). The following points summarize
the main conclusions in this work:
\begin{enumerate}

\item For a given value of wind magnetization, $\Upsilon$, (or a given
  value of $\Upsilon_{open}$), a hotter wind is faster, reaches the
  Alfv\'{e}n speed closer to the star and, as a consequence, the
  torque exerted on the surface of the star decreases.

\item We present two formulae that estimate the size of the
  torque-averaged Alfv\'{e}n radius: one that depends on parameter
  $\Upsilon$, which is based on stellar-surface parameters, and a
  second one that depends on $\Upsilon_{open}$, which is based on the
  amount of open magnetic flux. Each formulation gives a simple power
  law for each coronal temperature. By substituting equation
  (\ref{eq_rAvsUpsilon}) into equation (\ref{eq_rA}), the stellar
  angular-mometum-loss rate due to a magnetized wind is
  \begin{eqnarray}
  \label{eq_TorqueForm_a}
  \tau_w = K_s^2 \Omega_* \upsilon_{esc}^{-2 m_s} \dot{M}_w^{1 - 2m_s}
  R_*^{2+4m_s} B_*^{4m_s},
  \end{eqnarray}
  which is useful if the dipole field strength at the stellar surface
  is known. Similarly, by combining equations (\ref{eq_rA}) and
  (\ref{eq_RAvsYopen}), we have
  \begin{eqnarray}
  \label{eq_TorqueForm_c}
  \tau_w = K_o^2 \Omega_* \upsilon_{esc}^{-2 m_o} \dot{M}_w^{1 - 2m_o}
  R_*^{2-4m_o} \Phi_{open}^{4m_o},
  \end{eqnarray}
  which is useful if the amount of the total open magnetic flux is
  known. The above relations can be used for studies of the rotational
  evolution of cool stars, and predict the torque on stars with
  dipolar magnetic fields, that are slow rotators, and exhibit coronal
  winds with $\gamma=1.05$. Four different flow temperatures 
  were studied, and the values of fitting constants, $K_s,m_s,K_o,m_o$
  for each temperature, can be found in Table \ref{tab_results_b}. 

\item Using a simplified analysis (in \S \ref{sec_BrakingLaws}), we
  identified that the wind acceleration profile is a key factor that
  determines how the torque scales with parameter $\Upsilon$ or
  $\Upsilon_{open}$. We found (in Figures \ref{rAvsYopenmod} and
  \ref{rAvsYmod}) that by including the
  dimensionless velocity term, $\upsilon_{esc}/\bar{\mathrm{V}}_{R_A}$,
  ($\bar{\mathrm{V}}_{R_A}$ is the wind's mean speed at the Alfv\'{e}n
  surface), in each of the two torque formulae, all the simulation
  data collapse into a unique power law, independent of the flow
  temperature. In other words, we propose that a key term that
  needs to be included in stellar-torque prescriptions when one
  considers stars with different coronal conditions (and consequently
  different wind acceleration profiles) is the average wind speed at
  the Alfv\'{e}n surface, \textit{whatever heats and expands the
    outflow}. This conclusion should be independent of the actual wind
  temperature or details of how the wind is driven, since the angular
  momentum flux primarily depends on the flow velocity, mass density,
  and the magnetic field properties (see e.g., eqns \ref{eq_Jdot} and
  \ref{eq_Lambda}).

\item By considering a power-law dependence of
  $\bar{\mathrm{V}}_{R_A}$ (i.e. wind's mean speed at the Alfv\'{e}n
  surface) in $<R_A>/R_*$, the torque-averaged Alfv\'{e}n radius can
  be expressed with an analytic form (see eqs \ref{eq_rAvsYopenmod_b},
  \ref{eq_rAvsYmod_b}), for a well-aproximated (or known) wind
  acceleration profile. Equations (\ref{eq_rA}),
  (\ref{eq_rAvsYmod_b}), and (\ref{eq_rA}), (\ref{eq_rAvsYopenmod_b}),
  then yield respectively,
  \begin{eqnarray}
  \label{eq_TorqueForm_b}
  \tau_w =  {K_{l}^{8/(4+q)} \over K_q^{2/(4+q)}}
    \Omega_* \upsilon_{esc}^{-2/(4+q)} \dot{M}_w^{(2+q)/(4+q)}
    \nonumber \\ \times R_*^{(12+2q)/(4+q)} B_*^{4/(4+q)},
  \end{eqnarray}
  and
  \begin{eqnarray}
  \label{eq_TorqueForm_d}
  \tau_w =  {K_{c}^{4/(2+q)} \over K_q^{2/(2+q)}} 
    \Omega_* \upsilon_{esc}^{-2/(2+q)} \dot{M}_w^{q/(2+q)}
    \nonumber \\ \times R_*^{2q/(2+q)} \Phi_{open}^{4/(2+q)}.
  \end{eqnarray}
  These equations are sucessors to equations (\ref{eq_TorqueForm_a}),
  (\ref{eq_TorqueForm_c}), since they drop the dependence of magnetic
  braking on the flow temperature. Thus, equations
  (\ref{eq_TorqueForm_b}), and (\ref{eq_TorqueForm_d}) should predict
  stellar torques for any given coronal temperature, but require the
  wind acceleration profile to be known. The values of fitting
  constants $K_q,q$, that determine the acceleration of the outflow,
  for the temperatures examined in this study, can be found in table
  \ref{tab_results_b} (see also Appendix \ref{Appendix_Kq_q_vs_CsVesc}
  for predictions on the values of these fitting constants over a
  continuous range of temperartures), and the values of $K_c$, $K_l$
  exist in subsections \ref{subsec_1D_RAvsYopen} and
  \ref{subsec_1D_RAvsY}, respectively. 
\end{enumerate}

In order to give an example of how our formulation can be used, we
apply it to the solar case. In general, the torque exerted on the Sun
(or any star) is an integrated quantity, and its value depends on a
sum over the local values of the angular momentum flux (see eqn
\ref{eq_Jdot}). During the solar minimum the solar wind comprises two
components, a fast and a slow wind (see also \S
\ref{sec_introduction}). Our wind models do not produce a bimodal
outflow, and thus, we expect that our estimated solar torque should
lie somewhere in-between the torques predicted by our fastest
(i.e. with $c_s/\upsilon_{esc}=0.4$) and one of our slower wind models
(i.e. with $c_s/\upsilon_{esc}=0.25$). To calculate the solar-wind
torque, we will use the open-flux formula, given by equation
(\ref{eq_TorqueForm_c}), because the open magnetic flux is measured in
the solar wind by in situ spacecraft. Furthermore, previous studies
\citep{Reville:2015ab,Finley:2017aa} showed this formulation to be
independent of the field geometry at the surface (see also eqn
\ref{eq_rAvsYopen_b}). \citet{Smith:2003ab,Smith:2008aa} show that the
open flux at solar minimum is typically $\sim 7 \times 10^{22}
\textrm{Mx}$. In addition, by using $\Omega_{\odot}=2.87 \times
10^{-6} \textrm{rad s}^\textsuperscript{-1}$, $\dot{M}_{\odot}=2
\times 10^{-14} M_{\odot} \textrm{yr}^\textsuperscript{-1}$, and the
corresponding values of $K_o$, $m_o$ for $c_s/\upsilon_{esc}=0.25\
\textrm{and}\ 0.4$, equation (\ref{eq_TorqueForm_c}) yields an
angular-momentum-loss rate of $0.9 \times 10^{30}$ and $2.3 \times
10^{30} \textrm{erg}$, respectively. These values agree with the solar
braking rate found by \citet{Pizzo:1983aa}, which is $2.5 - 3.8 \times
10^{30} \textrm{erg}$, and that found by \citet{Li:1999aa}, which is
$2.1 \times 10^{30} \textrm{erg}$.

Even though we have used a simplified wind modeling (i.e. polytropic),
the proposed torque formalism should work for any cool star with a
known wind acceleration, mass-loss rate, and magnetic properties. The
physical mechanisms that expand flows from the hot coronae of cool
stars are still unknown \citep[e.g.][]{Cranmer:2012aa,Cranmer:2015aa},
but it is certain from early studies \citep[e.g.][]{Holzer:1977aa}
that the physics of coronal heating is more complex than simple
thermal-pressure expansion. The most modern ideas include
Alfv\'{e}n-wave dissipation
\citep[e.g.][]{Suzuki:2005aa,Cranmer:2007aa,Sokolov:2013aa,van-der-Holst:2014aa},
which work as an energy source and drive magnetized outflows. However
our full parameter space, with the range in flow temperatures that has 
been studied, should produce wind acceleration profiles within the
range that exist in real stars.

Future  work is needed to test the effects of more realistic wind
physics (e.g. with variations in the polytropic index $\gamma$ or
improved coronal heating models), and extending the study into the
fast-magnetic-rotator regime.

\acknowledgments
The authors thank Claudio Zanni, Victor R\'{e}ville, Sasha Brun,
Victor See, Adam Finley, and Matthew Gent for helpful discussions and
comments on the manuscript. G.P. acknowledges support from the
University of Exeter CEMPS through a Ph.D. studentship. This project
has received funding from the European Research Council (ERC) under
the European Union’s Horizon 2020 research and innovation programme
(grant AWESoMeStars, agreement No 682393). We also thank Andrea
Mignone and others for the development and maintenance of the PLUTO
code. This research has made use of NASA’s Astrophysics Data
System. All the figures within this work were produced using the
Python-library Matplotlib \citep{Hunter:2007aa}.

\appendix

\section{A. Periodic wind solutions} \label{Appendix_SteadyState}

\begin{figure}
\plotone{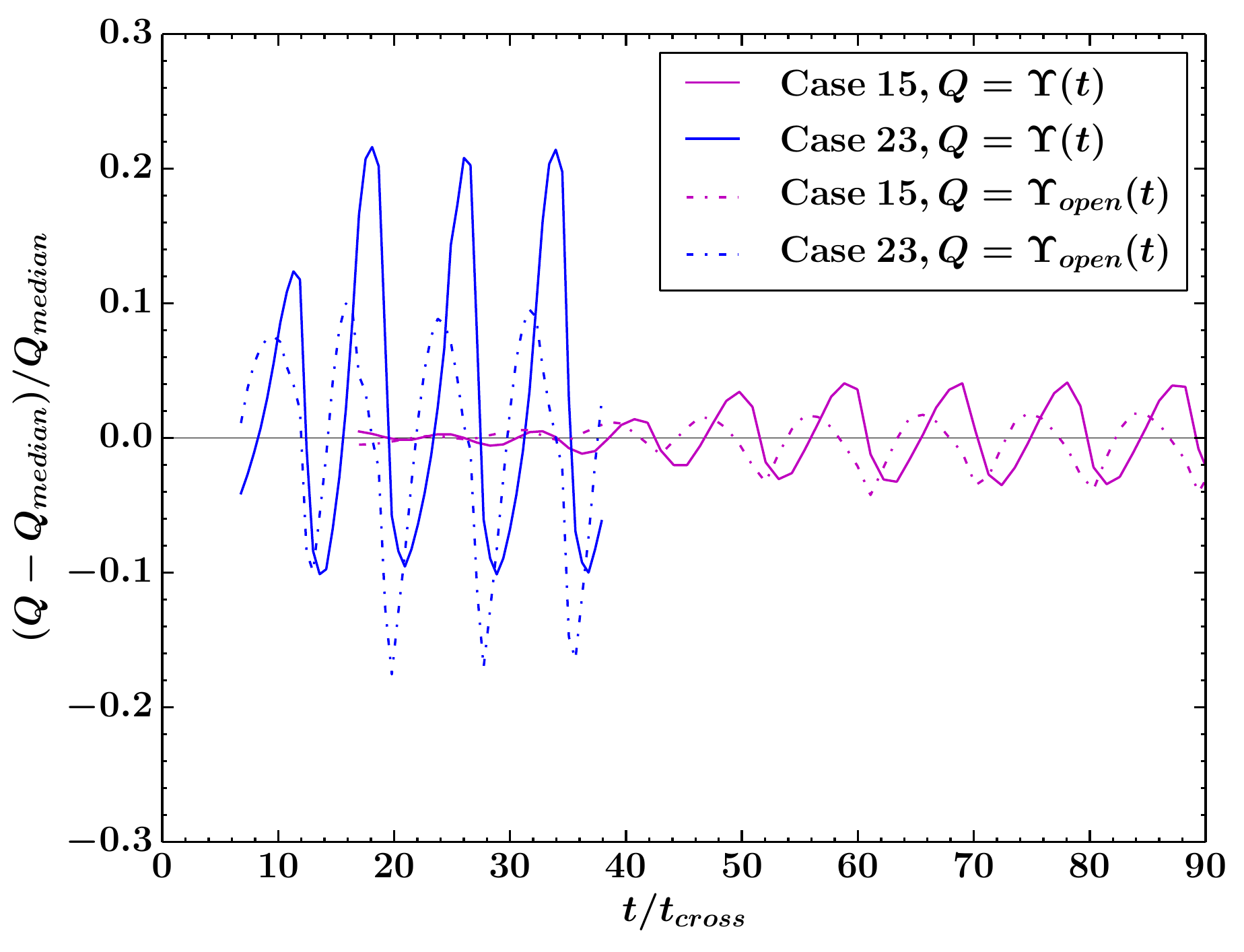}
\caption{Variations of $\Upsilon(t)$ and $\Upsilon_{open}(t)$ relative
  to median values of $\Upsilon$ and $\Upsilon_{open}$,
  respectively, versus number of crossing times $t/t_{cross}$. Two
  cases are shown, represented by the magenta lines (case 15) and the
  blue lines (case 23). The solid lines show the variations in
  parameter $\Upsilon$ and the dotted-dashed lines show the variations
  $\Upsilon_{open}$, respectively.}
\label{Y_Yopen}
\end{figure}

Each simulation is stopped when the solution relaxes to a steady
state. About half of our wind solutions show a steady nature to some
tolerance (see below), and the rest are periodic (or quasi-steady
state) due to magnetic reconnections (due to numerical diffusion) at
the neutral point (or cusp) located  at the equatorial region of each
simulation. As a consequence, a perfect steady-state solution cannot
be obtained. Similar features has been noted by \citet{Washimi:1993aa},
who found that the neutral point can have a non-steady behavior.

Due to this non-stationary nature of the equatorial region in some of
our simulations, the fluxes passing through spherical surfaces, within
our computational domain, are not constant in radius and
time. As a result, parameters $\Upsilon$, $\Upsilon_{open}$, and the
effective Alfv\'{e}n radius, $<R_A>/R_*$, show a dependence in both
radius and time (whereas they should be constant for an ideal and
steady-state MHD wind). However, the fluctuations of $\dot{M}_w$,
$\tau_w$, and $\Phi_{open}$, are well behaved and oscillatory, and the
amplitude of the oscillations is constant in both $r$ and $t$. In
order to derive single values for $\dot{M}_w$, $\tau_w$, and
$\Phi_{open}$, we used their median values in both $r$ (as dicsussed
in \S \ref{subsec_OutflowRates}), and $t$, where the value of a
quantity was taken to be its median value after the initial transient
phase of the simulation (i.e. typically after $\sim 10$ crossing
times). These global values of $\dot{M}_w$, $\tau_w$ and
$\Phi_{open}$, are then used to calculate $\Upsilon$,
$\Upsilon_{open}$, and $<R_A>R_*$, for each case.

The relative errors of the time-varying $\Upsilon(t)$ and
$\Upsilon_{open}(t)$ to the global values of $\Upsilon$ and
$\Upsilon_{open}$ are shown in figure \ref{Y_Yopen}, as a function of
number of wind crossing times, $t/t_{cross}$, (where
$t_{cross}=50R_*/\upsilon_{esc}$). The relative error of a given
quantity to its median value is taken to be $(Q -
Q_{median})/Q_{median}$, where $Q$ is $\Upsilon$ or
$\Upsilon_{open}$. Two cases are presented [i.e. case 15 (23) has the
magenta (blue) line]. The solid lines correspond to the relative
errors in $\Upsilon(t)$, and the dotted-dashed lines show the relative
errors in $\Upsilon_{open}(t)$. From figure \ref{Y_Yopen} it is clear,
that $\Upsilon(t)$ and $\Upsilon_{open}(t)$ fluctuate in time, and
furthermore are well-behaved functions of $t$. The variations in
$<R_A>/R_*$ are smaller in magnitude, compared to the variations seen
in $\Upsilon$ and $\Upsilon_{open}$, for a given wind solution. For
example, case 23, shown in figure \ref{Y_Yopen}, exhibit variations
in $<R_A>/R_*$ of about $2\%$ (compared to the range of variations in
$\Upsilon$ shown in the figure). 

Overall, for this study of 30 wind solutions, we obtained 16
steady-state wind solutions, meaning that the fluctuations in quantity
$\Upsilon$, are not noticeable or less than $2\%$. 7 wind solutions
show variations in the range between $2\%$ and $10\%$, and 
in 7  simulations the variations in $\Upsilon(t)$ are between $10\%$
and $30\%$. Additionally, we did not see any systematic difference in
the trends shown in this paper between the steady and periodic cases.

\section{B. Accuracy of the Numerical Solutions} \label{Appendix_Accuracy}

\begin{figure}
\plotone{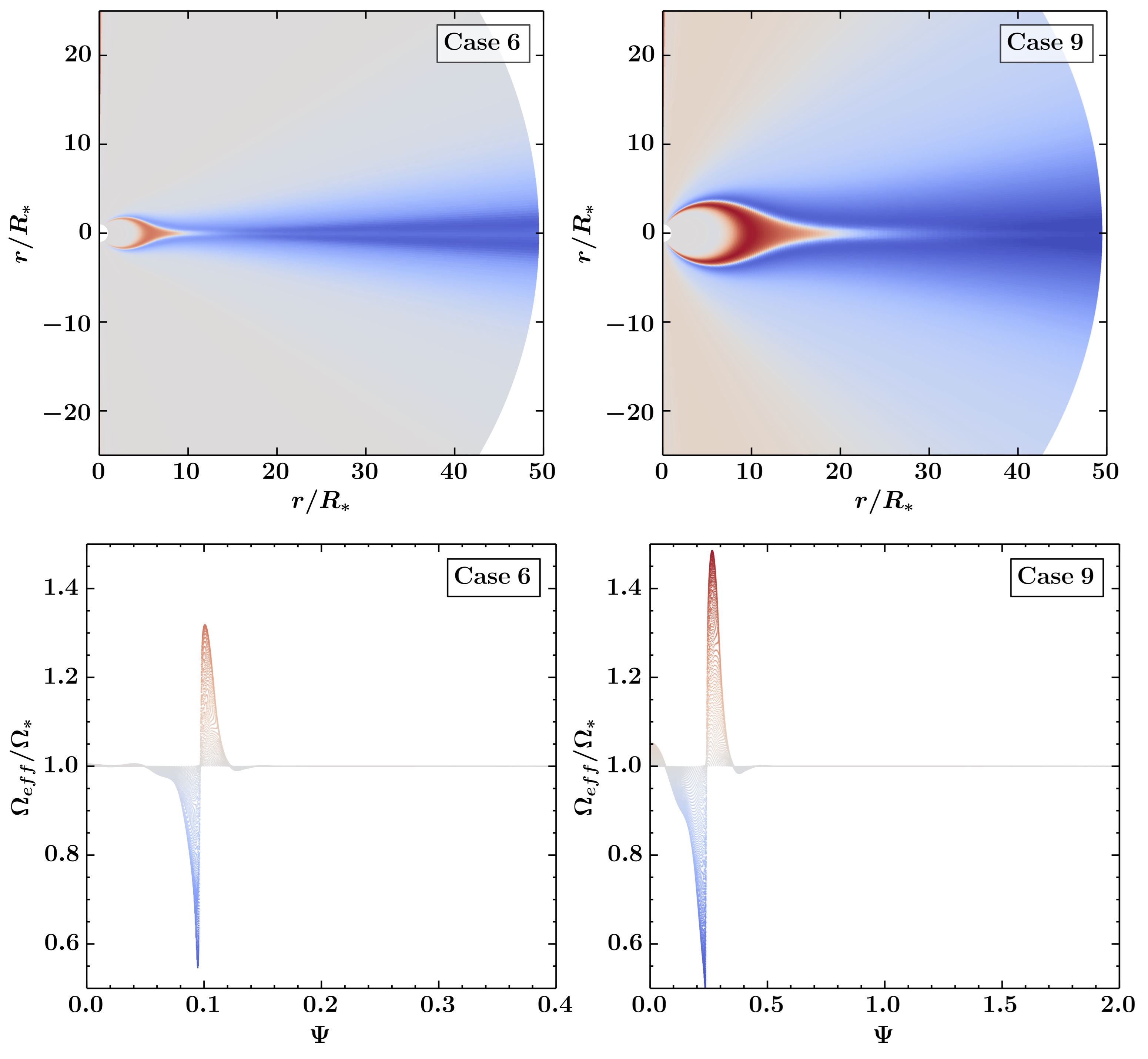}
\caption{Normalized effective rotation of field lines for two cases of
  the study. In the top panels, $\Omega_{eff}/\Omega_*$ is visualized
  as a 2D color-scale map. In the bottom panels,
  $\Omega_{eff}/\Omega_*$ is plotted versus the magnetic stream
  function $\Psi$. In the bottom plots, each plotted point represents
  a grid cell in the computational grid and each field line is
  associated with a unique value of $\Psi$. The color scale is the
  same for each plot. By design, the polar fieldline has a value of
  $\Psi=0$. The open-field region has a $\Psi$ that varies between 0
  and 0.1 for case 6 (bottom left panel), and between 0 and 0.25 for
  case 9 (bottom right panel).}
\label{OmegaEff}
\end{figure}

For ideal, axisymmetric, and steady-state, MHD outflows, there are
five scalar quantitities (i.e. derivative of the stream function or
mass flux per magnetic flux, Bernoulli or energy function, entropy,
specific angular momentum on a given stream function, effective
rotation rate of the field lines) that are constants of motion along
each field line \citep[e.g.][]{Heinemann:1978aa,Lovelace:1986aa,Ustyugova:1999aa,Keppens:2000aa}.
In order to examine the accuracy of each of our numerical solutions,
we check that each of the above quantities are conserved within some
tolerance. As shown by \citet{Zanni:2009aa} a difficult quantity to
conserve, and critical in order to measure accurate stellar torques,
is the effective rotation rate of the field lines,
$\Omega_{eff}$. Solving equation (\ref{eq_Vphi}) for $\Omega_*$, the
effective rotation of the field lines is defined as 
\begin{eqnarray}
\label{eq_OmegaEff}
\Omega_{eff}(\Psi) \equiv {1 \over {r \sin\theta}} \left({ \upsilon_{\phi}
  - {\upsilon_p \over B_p } B_{\phi}}\right), 
\end{eqnarray}
where $\Psi$ is the magnetic stream function, given in spherical
coordinates as $\Psi=r \sin\theta A$, where $r$ is the spherical
radius, $A$ is the scalar magnetic field potential
(i.e. $\bm{\mathrm{B}}_p = \nabla\times A \hat{\phi}$).
Each field line has a unique value of $\Psi$. Since the stream
function is a function of a scalar potential, $\Psi$ can be
determined everywhere by specifying its value at a single point. We
choose that $\Psi$ is zero at the pole, on the stellar surface
(i.e. $\Psi=0$ for $\theta=0$ and $r=R_*$), and as a result the first
polar field line will have a $\Psi$-value of zero.

In the ideal MHD regime, for any axisymmetric and steady-state wind
solution, equation (\ref{eq_Vphi}) should hold throughout the numerical
domain, and the plasma, which flows along the field lines, should
rotate such that the ratio $\Omega_{eff}/\Omega_*$ is equal to
unity. Any deviations from this value occur due to numerical diffusion
and non-stationary wind solutions. The crucial ingredients to achieve
correct rotation for the matter around the star are the boundary
conditions on $\upsilon_{\phi}$ and $B_{\phi}$, imposed on the inner
boundary (i.e. stellar surface) of the computational domain, as
pointed out in \citet{Zanni:2009aa}. For our simulations, the
torroidal speed of the plasma is enforced in the stellar boubdary via
equation (\ref{eq_Vphi}) and $B_{\phi}$ is linearly extrapolated
(i.e. $\partial B_{\phi}/ \partial r = \textrm{const.}$) into the ghost
zones, a boundary condition that works well for the current
stellar-wind numerical setup \citep[for a more detailed discussion on
different boundary conditions on $B_{\phi}$ see
also][]{Zanni:2009aa}.

In Figure \ref{OmegaEff}, the behavior of the normalized effective
rotation rate is presented as a 2D-color-scale plot (top panels), and
in a $\Omega_{eff}/\Omega_*$-versus-$\Psi$ plot (bottom panels), for
two numerical wind solutions of the study. The two cases shown are,
one that is typical  (case 6), and one (case 9) that exhibits among
the largest errors in the conservation of $\Omega_{eff}$. In the top
panels of figure \ref{OmegaEff}, the regions in the plots coloured
with grey correspond to an $\Omega_{eff}/\Omega_*$ that is equal to
unity. The blue and red regions correspond to
$\Omega_{eff}/\Omega_*<1$ and  $\Omega_{eff}/\Omega_*>1$,
respectively. For example, in case 6, we idenitfy that 
$\Omega_{eff}/\Omega_*$ is not conserved along field lines located at
mid-latitudes, adjacent to the dead zone, where steep gradients of
$\upsilon_{pol}$ and $B_{\phi}$ enhance the numerical diffusion. A
measure of how $\Omega_{eff}/\Omega_*$ deviates from unity, for these
two simulations, is given in the bottom panels of Figure
\ref{OmegaEff}. Each point in the bottom panels represent a grid cell,
within our domain, and every value of $\Psi$ correspond to a different
field line. Values of $\Psi$ from 0 to about 0.1 (case 6), and from 0
to about 0.25 (case 9) correspond to open field lines, in which the
wind flows outwards, and the rest of $\Psi$ values
represent closed magnetic loops. For case 6, we observe that
some open-field lines sub-rotate (up to $40\%$), and some closed-field
lines over-rotate (up to $30\%$). A comparison of
$\Omega_{eff}/\Omega_*$ between the two cases reveals that the errors
for case 9 (and cases with a high wind magnetization) are shifted to
the left because such simulations produce less fractional open
flux. For these cases the dead zones are more extended, cover most of
the stellar surafce, and as a result most of the open-field lines are
influenced by numerical errors. This can be easily seen in
top right panel in which the grey-shaded regions significantly
decrease compared to typical cases with median or low values of 
$\Upsilon$ (top left panel). Futhermore, the amplitude of the errors
becomes bigger in case 9, (see bottom left and right panel) as a
consequence of a wind that is more magnetized and due to this faster
(i.e. even steeper gradients of $\upsilon_{pol}$ and $B_{\phi}$). In
other words, numerical errors are more significant in simulations with
high wind magnetization. 

One way to reduce these non-ideal features is to increase the
resolution of the computational domain. For example, our resolution
studies (not shown) indicate that by doubling the number of cells in
the $\theta$ direction, numerical errors in $\Omega_{eff}/\Omega_*$
decrease, but the torque-averaged $<R_A>/R_*$, for most cases increase
only by a few precent. Bigger differences in the values of
$<R_A>/R_*$, due to a higher grid resolution, are observed in
simulations with $\Upsilon$ above $10^4$, but even for these cases
$<R_A>/R_*$ does not increase by more than $10\%$. These systematic
errors suggest that a more accurate numerical treatment would lead to
slightly steeper power laws in the trends shown in Figures \ref{rAvsY} -
\ref{rAvsYopenmod} and \ref{rAvsYmod}.

\section{C. Towards Predicting Torque for any
  Temperature} \label{Appendix_Kq_q_vs_CsVesc}

\begin{figure}
\plotone{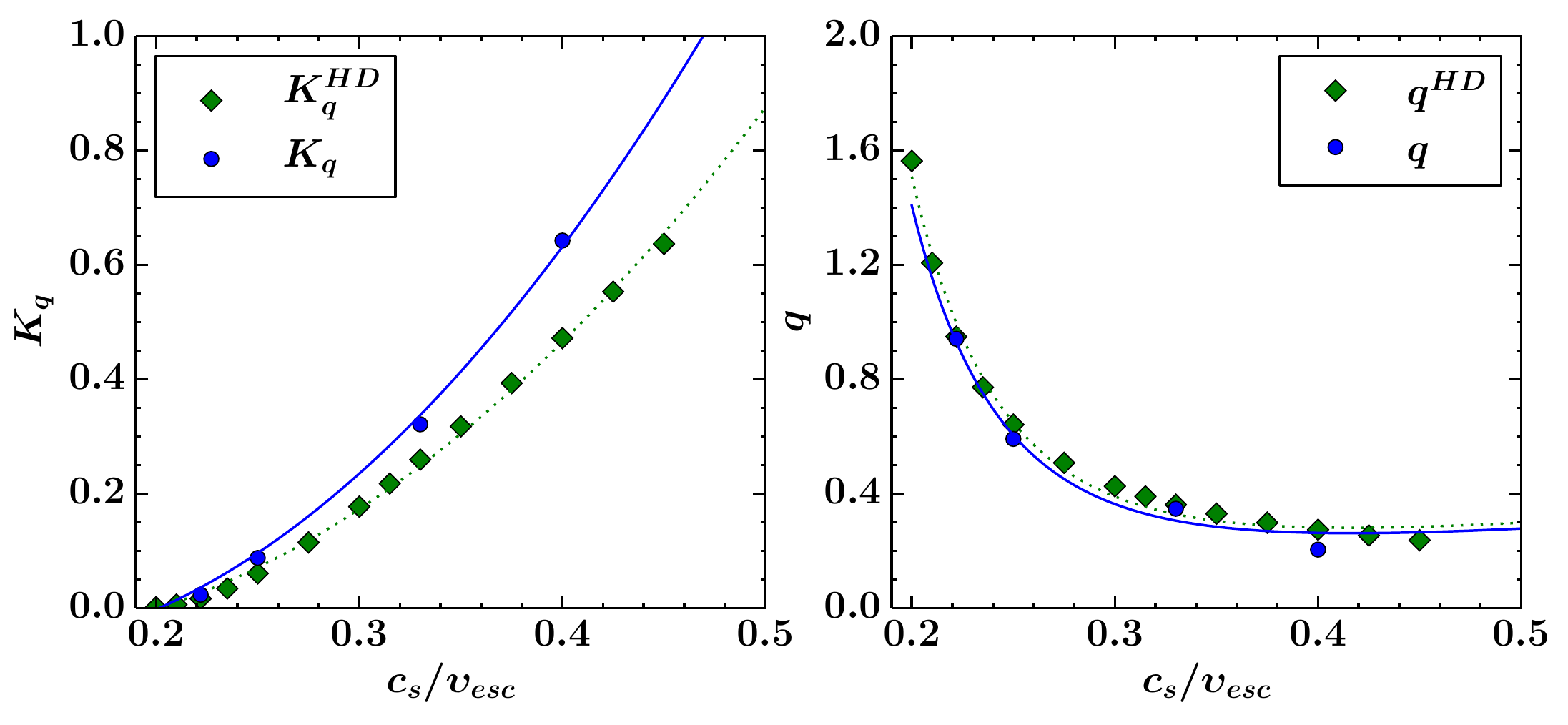}
\caption{Fitting constants $K_q$ (left panel), $q$ (right panel) of
  equation (\ref{eq_VrA}) versus parameter $c_s/\upsilon_{esc}$. The
  blue circles correspond to the values of $K_q$ and $q$ from the
  velocity laws presented in figure \ref{VrAvsRA}. The green diamonds
  correspond to the fitting constants $K_q^{HD}$ and $q^{HD}$, which
  have been obtained from 1D, HD wind speed profiles. The dotted lines
  fit the green data points, according to equations (\ref{eq_KqHD}) and
  (\ref{eq_qHD}). The blue solid lines show the fitting functions
  (eq. \ref{eq_Kq} and \ref{eq_q}) for $K_q$ and $q$, respectively.}
\label{Kq_q_CsVesc}
\end{figure}

In this Appendix we present empirical relations that predict the
fitting constants $K_q$, $q$, used to prescribe the wind speed at the 
Alfv\'{e}n radius (see eq. \ref{eq_VrA}), as functions of the input
parameter $c_s/\upsilon_{esc}$. $K_q$ and $q$ are needed in order to
estimate the torque-averaged Alfv\'{e}n radius (see
eqs. \ref{eq_rAvsYopenmod_b}, \ref{eq_rAvsYmod_b}), and since our
study investigated only four different flow temperatures, and their
corresponing wind acceleration profiles, our aim is to provide a
practical method that could give $K_q$ and $q$ over a larger, and
continuous range of $c_s/\upsilon_{esc}$. This method should work for
any continuous range of $c_s/\upsilon_{esc}$, for polytropic winds
with $\gamma=1.05$. It also suggests how to generalize for other winds,
but we do not test that here.

The values of $K_q$ and $q$ versus parameter $c_s/\upsilon_{esc}$ are
shown in Figure \ref{Kq_q_CsVesc} for our four temperatures (blue
circles). All the values of $q$ are positive, in the range between
zero and unity. There is no physical reasoning for not getting wind
solutions with values of $q$, such as $q>1$, but a $q=0$ is the lower
limit for any accelerating flow. Regardless of the obvious trends in
figure \ref{Kq_q_CsVesc}, (i.e. $K_q$ and $q$ monotonically increase
and decrease with an increasing $c_s/\upsilon_{esc}$, respectively),
any function that could possibly represent (or fit) these data points,
would be rather biased due to the small number of data points (only
four). Therefore, in order to construct functions that can fit the
data in figure \ref{Kq_q_CsVesc}, we employ the following approach. In
Figure \ref{VrAvsRA}, we demonstrated that the
$\bar{\mathrm{V}}_{R_A}/\upsilon_{esc}$-versus-$<R_A>/R_*$ data 
points exhibit a behavior similar to the shape of the radial-velocity
profiles (i.e. $\upsilon_{r}/\upsilon_{esc}$ versus $r/R_*$) of the 1D,
hydrodynamic, winds shown in Figures \ref{HDwind_log_Vr} and
\ref{VrAvsRA}. Based on that, one can infer what
$\bar{\mathrm{V}}_{R_A}$ would be for any given flow temperature (or
any given value of $c_s/\upsilon_{esc}$) from polytropic, Parker's
winds of that value of $c_s/\upsilon_{esc}$. Hence, we produce 14
Parker's wind models, in which parameter $c_s/\upsilon_{esc}$ varies
between 0.2 and 0.45. The velocity profiles of these winds are
functions of radial distance from the surface of the star. Then, we
treat any radial distance, of these profiles, as a potential
Alfv\'{e}n radius, and its corresponding flow velocity as the mean
speed of the outflow at the Alfv\'{e}n radius
(i.e. $\bar{\mathrm{V}}_{R_A}$). Following equation (\ref{eq_VrA}), we
fit these HD wind speed profiles, assuming that the flow speed is a
power law in radial distance (i.e. $\upsilon(r) \propto K_q^{HD}
r^{q^{HD}}$). Since for the entire study the minimum and maximun value
of $<R_A>/R_*$ is $3.27R_*$ and $25.3R_*$, respectively, the HD wind
speed profiles are fitted for radial distances that range between
$4R_*$ and $25R_*$. We obtain 14 new pairs of the dimensionless
fitting constants $K_q^{HD}$ and $q^{HD}$, also shown in figure
\ref{Kq_q_CsVesc} as green diamonds. The values of $K_q^{HD}$ and
$q^{HD}$ can be slightly influenced by considering a different range
in radii, in order to fit these HD wind speed profiles. The following
empirical functions can fit the new data points (i.e. $K_q^{HD}$ and
$q^{HD}$)
\begin{eqnarray}
\label{eq_KqHD}
K_q^{HD} = \alpha_1  (c_s/\upsilon_{esc})^2 + \beta_1  (c_s/\upsilon_{esc}),
\end{eqnarray}
\begin{eqnarray}
\label{eq_qHD}
q^{HD} = \alpha_2 (c_s/\upsilon_{esc})^{\nu_2} + \beta_2 (c_s/\upsilon_{esc}),
\end{eqnarray}
where $\alpha_1$, $\alpha_2$, $\beta_1$, $\beta_2$, and $\nu_2$ are
fitting coefficients. The best-fit values are $\alpha_1=5.87$,
$\beta_1=-1.18$, for equation (\ref{eq_KqHD}), and
$\alpha_2=0.000979$, $\nu_2=-4.51$, $\beta_2=0.553$, for equation 
(\ref{eq_qHD}). Equations (\ref{eq_KqHD}) and (\ref{eq_qHD}) are
represented in both panels of Figure \ref{Kq_q_CsVesc} by the
green dotted curves.

Equations (\ref{eq_KqHD}) and (\ref{eq_qHD}) can represent the 4 data
points (blue circles) in figure \ref{Kq_q_CsVesc}, just by including a
multiplicity factor. Indeed the blue, solid lines in figure
\ref{Kq_q_CsVesc} show that the data of $K_q$ and $q$ can be fitted by
functions in the form of 
\begin{eqnarray}
\label{eq_Kq}
K_q = D_1 [\alpha_1  (c_s/\upsilon_{esc})^2 + \beta_1  (c_s/\upsilon_{esc})],
\end{eqnarray}
\begin{eqnarray}
\label{eq_q}
q = D_2 [\alpha_2 (c_s/\upsilon_{esc})^{\nu_2} + \beta_2 (c_s/\upsilon_{esc})],
\end{eqnarray}
where again $D_1$, $D_2$ are fitting constants, and their best-fit
values are found to be is $D_1=1.36$, and $D_2=0.932$. In conclusion,
equations (\ref{eq_Kq}) and (\ref{eq_q}) can successfully predict the
values of dimensionless fitting constants $K_q$ and $q$ for any value
of parameter $c_s/\upsilon_{esc}$ in the range between 0.2 and 0.45,
for thermally-driven winds from slow-rotating stars, with dipolar
fields, and a fixed value of polytropic index equal to 1.05.

\section{D. Complete Grid of Simulations} \label{Appendix_Sims}

\begin{figure*}
\centering
%\plotone{figures/Grid_Sims.pdf}
\includegraphics[scale=0.05]{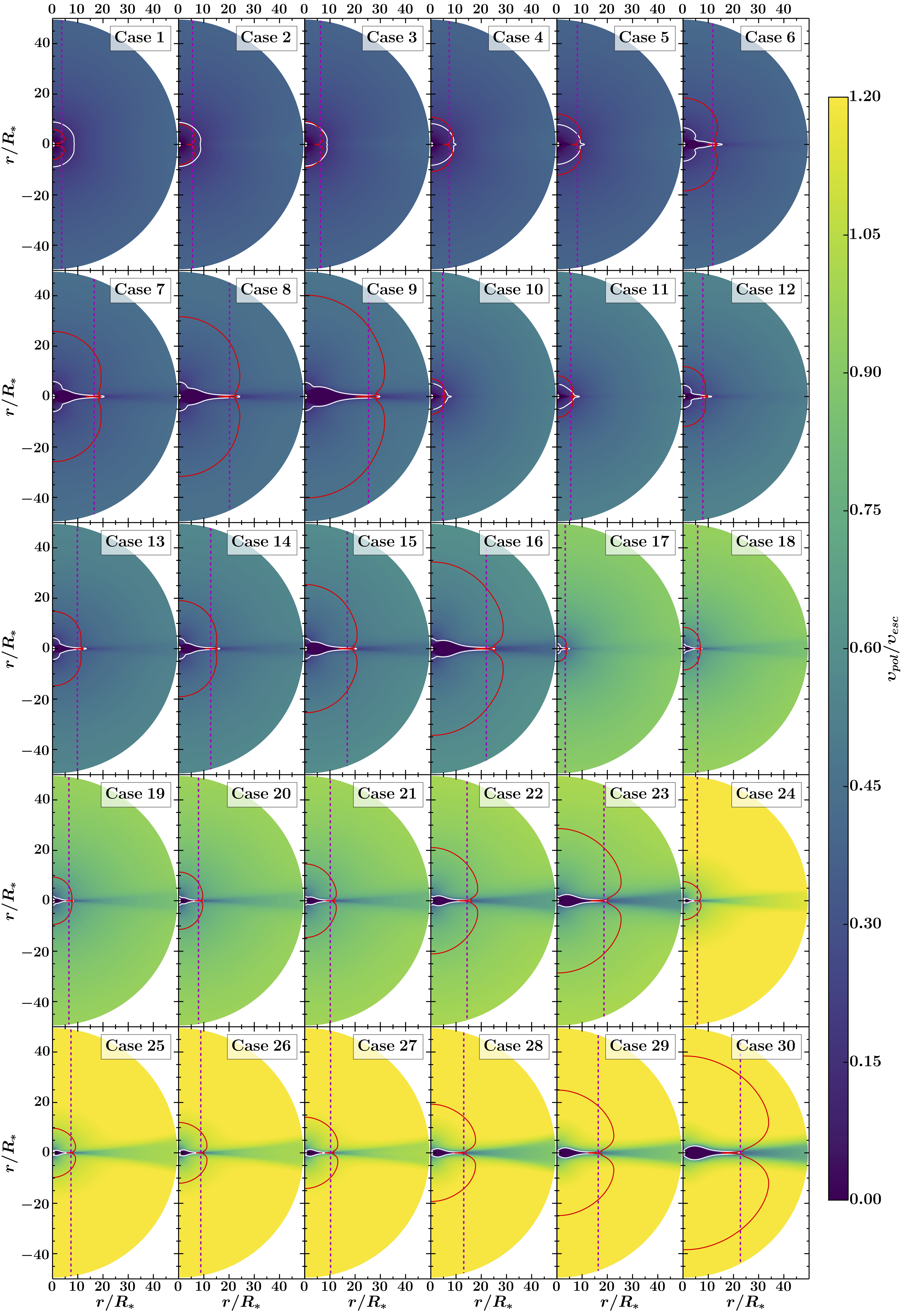}
\caption{Wind poloidal velocity colormaps of the entire study. White
  and red lines represent the sonic, and Alfv\'{e}n surfaces
  respectively. The magenta dashed lines show the location of
  torque-averaged Alfv\'{e}n radius (or effective lever
  arm). Simulations 1 to 9, 10 to 16, 17 to 23, 24 to 30, have
  respectively $c_s/\upsilon_{esc}=0.2219,0.25, 0.33, 0.4$.} 
\label{Grid_Sims}
\end{figure*}

Figure \ref{Grid_Sims} presents color-scale plots of the wind's
poloidal velocity, for all the numerical solutions of this
study. Cases 1 to 9, 10 to 16, 17 to 23, 24 to 30, have respectively
$c_s/\upsilon_{esc}=0.2219,0.25, 0.33, 0.4$. Each panel in figure
\ref{Grid_Sims} shows the full computational grid, the location and
the shape of the wind's critical surfaces. The sonic and Alfv\'{e}n
surfaces are depicted with white and red solid lines, respectively.
The magenta dashed lines show the effective lever arm,
$<R_A>/R_*$, that brakes the stellar rotation. A different coronal
temperature (primarily) and a higher wind magnetization (to a lesser
extent) affects the outflow speed and acceleration profile. This
feature can be seen by the changes in the color scale of each
panel. Overall, for a given value of the wind magnetization,
$\Upsilon$, a hotter wind reaches the Alfv\'{e}n surface in a shorter
disctance from the stellar surface, the size of the lever arm
decreases, and as a result magnetic torque is reduced.


\begin{thebibliography}{}
%\expandafter\ifx\csname natexlab\endcsname\relax\def\natexlab#1{#1}\fi

\bibitem[{{Alvarado-G{\'o}mez} {et~al.}(2016){Alvarado-G{\'o}mez}, {Hussain},
  {Cohen}, {Drake}, {Garraffo}, {Grunhut}, \&
  {Gombosi}}]{Alvarado-Gomez:2016ab}
{Alvarado-G{\'o}mez}, J.~D., {Hussain}, G.~A.~J., {Cohen}, O., {et~al.} 2016,
  \aap, 594, A95

\bibitem[{{Amard} {et~al.}(2016){Amard}, {Palacios}, {Charbonnel}, {Gallet}, \&
  {Bouvier}}]{Amard:2016aa}
{Amard}, L., {Palacios}, A., {Charbonnel}, C., {Gallet}, F., \& {Bouvier}, J.
  2016, \aap, 587, A105

\bibitem[{{Aschwanden}(2005)}]{Aschwanden:2005aa}
{Aschwanden}, M.~J. 2005, {Physics of the Solar Corona. An Introduction with
  Problems and Solutions (2nd edition)} (Springer-Verlag Berlin Heidelberg)

\bibitem[{{Balsara} \& {Spicer}(1999)}]{Balsara:1999aa}
{Balsara}, D.~S., \& {Spicer}, D.~S. 1999, Journal of Computational Physics,
  149, 270

\bibitem[{{Baraffe} {et~al.}(1998){Baraffe}, {Chabrier}, {Allard}, \&
  {Hauschildt}}]{Baraffe:1998aa}
{Baraffe}, I., {Chabrier}, G., {Allard}, F., \& {Hauschildt}, P.~H. 1998, \aap,
  337, 403

\bibitem[{{Barnes}(2003)}]{Barnes:2003aa}
{Barnes}, S.~A. 2003, \apj, 586, 464

\bibitem[{{Barnes}(2010)}]{Barnes:2010aa}
---. 2010, \apj, 722, 222

\bibitem[{{Belcher} \& {MacGregor}(1976)}]{Belcher:1976aa}
{Belcher}, J.~W., \& {MacGregor}, K.~B. 1976, \apj, 210, 498

\bibitem[{{Bouvier} {et~al.}(2014){Bouvier}, {Matt}, {Mohanty}, {Scholz},
  {Stassun}, \& {Zanni}}]{Bouvier:2014aa}
{Bouvier}, J., {Matt}, S.~P., {Mohanty}, S., {et~al.} 2014, Protostars and
  Planets VI, 433

\bibitem[{{Cohen} \& {Drake}(2014)}]{Cohen:2014aa}
{Cohen}, O., \& {Drake}, J.~J. 2014, \apj, 783, 55

\bibitem[{{Cohen} {et~al.}(2007){Cohen}, {Sokolov}, {Roussev}, {Arge},
  {Manchester}, {Gombosi}, {Frazin}, {Park}, {Butala}, {Kamalabadi}, \&
  {Velli}}]{Cohen:2007aa}
{Cohen}, O., {Sokolov}, I.~V., {Roussev}, I.~I., {et~al.} 2007, \apjl, 654,
  L163

\bibitem[{{Cranmer}(2012)}]{Cranmer:2012aa}
{Cranmer}, S.~R. 2012, \ssr, 172, 145

\bibitem[{{Cranmer} {et~al.}(2015){Cranmer}, {Asgari-Targhi}, {Miralles},
  {Raymond}, {Strachan}, {Tian}, \& {Woolsey}}]{Cranmer:2015aa}
{Cranmer}, S.~R., {Asgari-Targhi}, M., {Miralles}, M.~P., {et~al.} 2015,
  Philosophical Transactions of the Royal Society of London Series A, 373,
  20140148

\bibitem[{{Cranmer} \& {Saar}(2011)}]{Cranmer:2011aa}
{Cranmer}, S.~R., \& {Saar}, S.~H. 2011, \apj, 741, 54

\bibitem[{{Cranmer} {et~al.}(2007){Cranmer}, {van Ballegooijen}, \&
  {Edgar}}]{Cranmer:2007aa}
{Cranmer}, S.~R., {van Ballegooijen}, A.~A., \& {Edgar}, R.~J. 2007, \apjs,
  171, 520

\bibitem[{{De Moortel} \& {Browning}(2015)}]{De-Moortel:2015aa}
{De Moortel}, I., \& {Browning}, P. 2015, Philosophical Transactions of the
  Royal Society of London Series A, 373, 20140269

\bibitem[{{Donati} \& {Brown}(1997)}]{Donati:1997aa}
{Donati}, J.-F., \& {Brown}, S.~F. 1997, \aap, 326, 1135

\bibitem[{{Donati} \& {Landstreet}(2009)}]{Donati:2009aa}
{Donati}, J.-F., \& {Landstreet}, J.~D. 2009, \araa, 47, 333

\bibitem[{{Finley} \& {Matt}(2017)}]{Finley:2017aa}
{Finley}, A.~J., \& {Matt}, S.~P. 2017, \apj, 845, 46

\bibitem[{{Fisk}(2003)}]{Fisk:2003aa}
{Fisk}, L.~A. 2003, Journal of Geophysical Research (Space Physics), 108, 1157

\bibitem[{{Gallet} \& {Bouvier}(2013)}]{Gallet:2013aa}
{Gallet}, F., \& {Bouvier}, J. 2013, \aap, 556, A36

\bibitem[{{Gallet} \& {Bouvier}(2015)}]{Gallet:2015aa}
---. 2015, \aap, 577, A98

\bibitem[{{Garraffo} {et~al.}(2015){Garraffo}, {Drake}, \&
  {Cohen}}]{Garraffo:2015ab}
{Garraffo}, C., {Drake}, J.~J., \& {Cohen}, O. 2015, \apjl, 807, L6

\bibitem[{{Garraffo} {et~al.}(2016){Garraffo}, {Drake}, \&
  {Cohen}}]{Garraffo:2016aa}
---. 2016, \aap, 595, A110

\bibitem[{{Hansteen} \& {Velli}(2012)}]{Hansteen:2012aa}
{Hansteen}, V.~H., \& {Velli}, M. 2012, \ssr, 172, 89

\bibitem[{{Heinemann} \& {Olbert}(1978)}]{Heinemann:1978aa}
{Heinemann}, M., \& {Olbert}, S. 1978, \jgr, 83, 2457

\bibitem[{{Holzer}(1977)}]{Holzer:1977aa}
{Holzer}, T.~E. 1977, \jgr, 82, 23

\bibitem[{{Holzwarth} \& {Jardine}(2007)}]{Holzwarth:2007aa}
{Holzwarth}, V., \& {Jardine}, M. 2007, \aap, 463, 11

\bibitem[{{Hunter}(2007)}]{Hunter:2007aa}
{Hunter}, J.~D. 2007, {Computing In Science \& Engineering}, 9, 90

\bibitem[{{Irwin} \& {Bouvier}(2009)}]{Irwin:2009aa}
{Irwin}, J., \& {Bouvier}, J. 2009, in IAU Symposium, Vol. 258, The Ages of
  Stars, ed. E.~E. {Mamajek}, D.~R. {Soderblom}, \& R.~F.~G. {Wyse}, 363--374

\bibitem[{{Johnstone} {et~al.}(2015{\natexlab{a}}){Johnstone}, {G{\"u}del},
  {Brott}, \& {L{\"u}ftinger}}]{Johnstone:2015ad}
{Johnstone}, C.~P., {G{\"u}del}, M., {Brott}, I., \& {L{\"u}ftinger}, T.
  2015{\natexlab{a}}, \aap, 577, A28

\bibitem[{{Johnstone} {et~al.}(2015{\natexlab{b}}){Johnstone}, {G{\"u}del},
  {L{\"u}ftinger}, {Toth}, \& {Brott}}]{Johnstone:2015aa}
{Johnstone}, C.~P., {G{\"u}del}, M., {L{\"u}ftinger}, T., {Toth}, G., \&
  {Brott}, I. 2015{\natexlab{b}}, \aap, 577, A27

\bibitem[{{Kawaler}(1988)}]{Kawaler:1988aa}
{Kawaler}, S.~D. 1988, \apj, 333, 236

\bibitem[{{Keppens} \& {Goedbloed}(1999)}]{Keppens:1999aa}
{Keppens}, R., \& {Goedbloed}, J.~P. 1999, \aap, 343, 251

\bibitem[{{Keppens} \& {Goedbloed}(2000)}]{Keppens:2000aa}
---. 2000, \apj, 530, 1036

\bibitem[{{Klimchuk}(2015)}]{Klimchuk:2015aa}
{Klimchuk}, J.~A. 2015, Philosophical Transactions of the Royal Society of
  London Series A, 373, 20140256

\bibitem[{{Kopp} \& {Holzer}(1976)}]{Kopp:1976aa}
{Kopp}, R.~A., \& {Holzer}, T.~E. 1976, \solphys, 49, 43

\bibitem[{{Kraft}(1967)}]{Kraft:1967aa}
{Kraft}, R.~P. 1967, \apj, 150, 551

\bibitem[{{Lamers} \& {Cassinelli}(1999)}]{Lamers:1999aa}
{Lamers}, H.~J.~G.~L.~M., \& {Cassinelli}, J.~P. 1999, {Introduction to Stellar
  Winds} (Cambridge, UK: Cambridge University Press), 452

\bibitem[{{Leer} \& {Holzer}(1980)}]{Leer:1980aa}
{Leer}, E., \& {Holzer}, T.~E. 1980, \jgr, 85, 4681

\bibitem[{{Li}(1999)}]{Li:1999aa}
{Li}, J. 1999, \mnras, 302, 203

\bibitem[{{Lovelace} {et~al.}(1986){Lovelace}, {Mehanian}, {Mobarry}, \&
  {Sulkanen}}]{Lovelace:1986aa}
{Lovelace}, R.~V.~E., {Mehanian}, C., {Mobarry}, C.~M., \& {Sulkanen}, M.~E.
  1986, \apjs, 62, 1

\bibitem[{{Low} \& {Tsinganos}(1986)}]{Low:1986aa}
{Low}, B.~C., \& {Tsinganos}, K. 1986, \apj, 302, 163

\bibitem[{{L{\"u}ftinger} {et~al.}(2015){L{\"u}ftinger}, {Vidotto}, \&
  {Johnstone}}]{Luftinger:2015aa}
{L{\"u}ftinger}, T., {Vidotto}, A.~A., \& {Johnstone}, C.~P. 2015, in
  Astrophysics and Space Science Library, Vol. 411, Characterizing Stellar and
  Exoplanetary Environments, ed. H.~{Lammer} \& M.~{Khodachenko}, 37

\bibitem[{{Matt} \& {Pudritz}(2008)}]{Matt:2008aa}
{Matt}, S., \& {Pudritz}, R.~E. 2008, \apj, 678, 1109

\bibitem[{{Matt} {et~al.}(2015){Matt}, {Brun}, {Baraffe}, {Bouvier}, \&
  {Chabrier}}]{Matt:2015aa}
{Matt}, S.~P., {Brun}, A.~S., {Baraffe}, I., {Bouvier}, J., \& {Chabrier}, G.
  2015, \apjl, 799, L23

\bibitem[{{Matt} {et~al.}(2012){Matt}, {MacGregor}, {Pinsonneault}, \&
  {Greene}}]{Matt:2012ab}
{Matt}, S.~P., {MacGregor}, K.~B., {Pinsonneault}, M.~H., \& {Greene}, T.~P.
  2012, \apjl, 754, L26

\bibitem[{{McComas} {et~al.}(2008){McComas}, {Ebert}, {Elliott}, {Goldstein},
  {Gosling}, {Schwadron}, \& {Skoug}}]{McComas:2008aa}
{McComas}, D.~J., {Ebert}, R.~W., {Elliott}, H.~A., {et~al.} 2008, \grl, 35,
  L18103

\bibitem[{{McComas} {et~al.}(2003){McComas}, {Elliott}, {Schwadron}, {Gosling},
  {Skoug}, \& {Goldstein}}]{McComas:2003aa}
{McComas}, D.~J., {Elliott}, H.~A., {Schwadron}, N.~A., {et~al.} 2003, \grl,
  30, 24

\bibitem[{{McComas} {et~al.}(2007){McComas}, {Velli}, {Lewis}, {Acton},
  {Balat-Pichelin}, {Bothmer}, {Dirling}, {Feldman}, {Gloeckler}, {Habbal},
  {Hassler}, {Mann}, {Matthaeus}, {McNutt}, {Mewaldt}, {Murphy}, {Ofman},
  {Sittler}, {Smith}, \& {Zurbuchen}}]{McComas:2007aa}
{McComas}, D.~J., {Velli}, M., {Lewis}, W.~S., {et~al.} 2007, Reviews of
  Geophysics, 45, RG1004

\bibitem[{{Meibom} {et~al.}(2015){Meibom}, {Barnes}, {Platais}, {Gilliland},
  {Latham}, \& {Mathieu}}]{Meibom:2015aa}
{Meibom}, S., {Barnes}, S.~A., {Platais}, I., {et~al.} 2015, \nat, 517, 589

\bibitem[{{Meibom} {et~al.}(2011){Meibom}, {Barnes}, {Latham}, {Batalha},
  {Borucki}, {Koch}, {Basri}, {Walkowicz}, {Janes}, {Jenkins}, {Van Cleve},
  {Haas}, {Bryson}, {Dupree}, {Furesz}, {Szentgyorgyi}, {Buchhave}, {Clarke},
  {Twicken}, \& {Quintana}}]{Meibom:2011aa}
{Meibom}, S., {Barnes}, S.~A., {Latham}, D.~W., {et~al.} 2011, \apjl, 733, L9

\bibitem[{{Mestel}(1968)}]{Mestel:1968aa}
{Mestel}, L. 1968, \mnras, 138, 359

\bibitem[{{Mestel}(1984)}]{Mestel:1984aa}
{Mestel}, L. 1984, in Lecture Notes in Physics, Berlin Springer Verlag, Vol.
  193, Cool Stars, Stellar Systems, and the Sun, ed. S.~L. {Baliunas} \&
  L.~{Hartmann}, 49

\bibitem[{{Mestel}(1999)}]{Mestel:1999aa}
---. 1999, {Stellar magnetism} (New York: Oxford University Press)

\bibitem[{{Mestel} \& {Spruit}(1987)}]{Mestel:1987aa}
{Mestel}, L., \& {Spruit}, H.~C. 1987, \mnras, 226, 57

\bibitem[{{Mignone} {et~al.}(2007){Mignone}, {Bodo}, {Massaglia}, {Matsakos},
  {Tesileanu}, {Zanni}, \& {Ferrari}}]{Mignone:2007aa}
{Mignone}, A., {Bodo}, G., {Massaglia}, S., {et~al.} 2007, \apjs, 170, 228

\bibitem[{{Miki{\'c}} {et~al.}(1999){Miki{\'c}}, {Linker}, {Schnack},
  {Lionello}, \& {Tarditi}}]{Mikic:1999aa}
{Miki{\'c}}, Z., {Linker}, J.~A., {Schnack}, D.~D., {Lionello}, R., \&
  {Tarditi}, A. 1999, Physics of Plasmas, 6, 2217

\bibitem[{{Ofman}(2004)}]{Ofman:2004aa}
{Ofman}, L. 2004, Advances in Space Research, 33, 681

\bibitem[{{Ofman}(2010)}]{Ofman:2010aa}
---. 2010, Living Reviews in Solar Physics, 7, 4

\bibitem[{{Owocki}(2009)}]{Owocki:2009aa}
{Owocki}, S. 2009, in EAS Publications Series, Vol.~39, EAS Publications
  Series, ed. C.~{Neiner} \& J.-P. {Zahn}, 223--254

\bibitem[{{Parker}(1958)}]{Parker:1958aa}
{Parker}, E.~N. 1958, \apj, 128, 664

\bibitem[{{Parker}(1963)}]{Parker:1963aa}
---. 1963, Interplanetary Dynamical Processes (New York: Interscience
  Publishers)

\bibitem[{{Pizzo} {et~al.}(1983){Pizzo}, {Schwenn}, {Marsch}, {Rosenbauer},
  {Muehlhaeuser}, \& {Neubauer}}]{Pizzo:1983aa}
{Pizzo}, V., {Schwenn}, R., {Marsch}, E., {et~al.} 1983, \apj, 271, 335

\bibitem[{{Pizzolato} {et~al.}(2003){Pizzolato}, {Maggio}, {Micela},
  {Sciortino}, \& {Ventura}}]{Pizzolato:2003aa}
{Pizzolato}, N., {Maggio}, A., {Micela}, G., {Sciortino}, S., \& {Ventura}, P.
  2003, \aap, 397, 147

\bibitem[{{Pneuman}(1966)}]{Pneuman:1966aa}
{Pneuman}, G.~W. 1966, \apj, 145, 242

\bibitem[{{Pneuman} \& {Kopp}(1971)}]{Pneuman:1971aa}
{Pneuman}, G.~W., \& {Kopp}, R.~A. 1971, \solphys, 18, 258

\bibitem[{Powell {et~al.}(1999)Powell, Roe, Linde, Gombosi, \&
  Zeeuw}]{Powell1999284}
Powell, K.~G., Roe, P.~L., Linde, T.~J., Gombosi, T.~I., \& Zeeuw, D. L.~D.
  1999, Journal of Computational Physics, 154, 284

\bibitem[{{Priest}(2014)}]{Priest:2014aa}
{Priest}, E. 2014, {Magnetohydrodynamics of the Sun} (Cambridge, UK: Cambridge
  University Press)

\bibitem[{{Reiners} \& {Mohanty}(2012)}]{Reiners:2012aa}
{Reiners}, A., \& {Mohanty}, S. 2012, \apj, 746, 43

\bibitem[{{R{\'e}ville} {et~al.}(2015{\natexlab{a}}){R{\'e}ville}, {Brun},
  {Matt}, {Strugarek}, \& {Pinto}}]{Reville:2015ab}
{R{\'e}ville}, V., {Brun}, A.~S., {Matt}, S.~P., {Strugarek}, A., \& {Pinto},
  R.~F. 2015{\natexlab{a}}, \apj, 798, 116

\bibitem[{{R{\'e}ville} {et~al.}(2015{\natexlab{b}}){R{\'e}ville}, {Brun},
  {Strugarek}, {Matt}, {Bouvier}, {Folsom}, \& {Petit}}]{Reville:2015aa}
{R{\'e}ville}, V., {Brun}, A.~S., {Strugarek}, A., {et~al.} 2015{\natexlab{b}},
  \apj, 814, 99

\bibitem[{{R{\'e}ville} {et~al.}(2016{\natexlab{a}}){R{\'e}ville}, {Folsom},
  {Strugarek}, \& {Brun}}]{Reville:2016aa}
{R{\'e}ville}, V., {Folsom}, C.~P., {Strugarek}, A., \& {Brun}, A.~S.
  2016{\natexlab{a}}, \apj, 832, 145

\bibitem[{{R{\'e}ville} {et~al.}(2016{\natexlab{b}}){R{\'e}ville}, {Folsom},
  {Strugarek}, \& {Brun}}]{Reville:2016ab}
{R{\'e}ville}, V., {Folsom}, C.~P., {Strugarek}, A., \& {Brun}, A.~S.
  2016{\natexlab{b}}, in 19th Cambridge Workshop on Cool Stars, Stellar
  Systems, and the Sun (CS19), 33

\bibitem[{{Riley} {et~al.}(2006){Riley}, {Linker}, {Miki{\'c}}, {Lionello},
  {Ledvina}, \& {Luhmann}}]{Riley:2006aa}
{Riley}, P., {Linker}, J.~A., {Miki{\'c}}, Z., {et~al.} 2006, \apj, 653, 1510

\bibitem[{{Sakurai}(1985)}]{Sakurai:1985aa}
{Sakurai}, T. 1985, \aap, 152, 121

\bibitem[{{Schatzman}(1962)}]{Schatzman:1962aa}
{Schatzman}, E. 1962, Annales d'Astrophysique, 25, 18

\bibitem[{{Schwadron} \& {McComas}(2003)}]{Schwadron:2003aa}
{Schwadron}, N.~A., \& {McComas}, D.~J. 2003, \apj, 599, 1395

\bibitem[{{Schwadron} \& {McComas}(2008)}]{Schwadron:2008aa}
---. 2008, \apjl, 686, L33

\bibitem[{{See} {et~al.}(2015){See}, {Jardine}, {Vidotto}, {Donati}, {Folsom},
  {Boro Saikia}, {Bouvier}, {Fares}, {Gregory}, {Hussain}, {Jeffers},
  {Marsden}, {Morin}, {Moutou}, {do Nascimento}, {Petit}, {Ros{\'e}n}, \&
  {Waite}}]{See:2015aa}
{See}, V., {Jardine}, M., {Vidotto}, A.~A., {et~al.} 2015, \mnras, 453, 4301

\bibitem[{{See} {et~al.}(2017){See}, {Jardine}, {Vidotto}, {Donati}, {Boro
  Saikia}, {Fares}, {Folsom}, {H{\'e}brard}, {Jeffers}, {Marsden}, {Morin},
  {Petit}, {Waite}, \& {BCool Collaboration}}]{See:2017aa}
---. 2017, \mnras, 466, 1542

\bibitem[{{Skumanich}(1972)}]{Skumanich:1972aa}
{Skumanich}, A. 1972, \apj, 171, 565

\bibitem[{{Smith} \& {Balogh}(2003)}]{Smith:2003ab}
{Smith}, E.~J., \& {Balogh}, A. 2003, in American Institute of Physics
  Conference Series, Vol. 679, Solar Wind Ten, ed. M.~{Velli}, R.~{Bruno},
  F.~{Malara}, \& B.~{Bucci}, 67--70

\bibitem[{{Smith} \& {Balogh}(2008)}]{Smith:2008aa}
{Smith}, E.~J., \& {Balogh}, A. 2008, \grl, 35, L22103

\bibitem[{{Sokolov} {et~al.}(2013){Sokolov}, {van der Holst}, {Oran}, {Downs},
  {Roussev}, {Jin}, {Manchester}, {Evans}, \& {Gombosi}}]{Sokolov:2013aa}
{Sokolov}, I.~V., {van der Holst}, B., {Oran}, R., {et~al.} 2013, \apj, 764, 23

\bibitem[{{Suzuki} {et~al.}(2013){Suzuki}, {Imada}, {Kataoka}, {Kato},
  {Matsumoto}, {Miyahara}, \& {Tsuneta}}]{Suzuki:2013aa}
{Suzuki}, T.~K., {Imada}, S., {Kataoka}, R., {et~al.} 2013, \pasj, 65, 98

\bibitem[{{Suzuki} \& {Inutsuka}(2005)}]{Suzuki:2005aa}
{Suzuki}, T.~K., \& {Inutsuka}, S.-i. 2005, \apjl, 632, L49

\bibitem[{{Testa} {et~al.}(2015){Testa}, {Saar}, \& {Drake}}]{Testa:2015aa}
{Testa}, P., {Saar}, S.~H., \& {Drake}, J.~J. 2015, Philosophical Transactions
  of the Royal Society of London Series A, 373, 20140259

\bibitem[{{Toro}(2009)}]{Toro:2009aa}
{Toro}, E.~F. 2009, {Riemann Solvers and Numerical Methods for Fluid Dynamics}
  (Springer-Verlag: Berlin)

\bibitem[{{Tout} \& {Pringle}(1992)}]{Tout:1992aa}
{Tout}, C.~A., \& {Pringle}, J.~E. 1992, \mnras, 256, 269

\bibitem[{{ud-Doula} \& {Owocki}(2002)}]{ud-Doula:2002aa}
{ud-Doula}, A., \& {Owocki}, S.~P. 2002, \apj, 576, 413

\bibitem[{{Ud-Doula} {et~al.}(2009){Ud-Doula}, {Owocki}, \&
  {Townsend}}]{ud-Doula:2009aa}
{Ud-Doula}, A., {Owocki}, S.~P., \& {Townsend}, R.~H.~D. 2009, \mnras, 392,
  1022

\bibitem[{{Ustyugova} {et~al.}(1999){Ustyugova}, {Koldoba}, {Romanova},
  {Chechetkin}, \& {Lovelace}}]{Ustyugova:1999aa}
{Ustyugova}, G.~V., {Koldoba}, A.~V., {Romanova}, M.~M., {Chechetkin}, V.~M.,
  \& {Lovelace}, R.~V.~E. 1999, \apj, 516, 221

\bibitem[{{van der Holst} {et~al.}(2014){van der Holst}, {Sokolov}, {Meng},
  {Jin}, {Manchester}, {T{\'o}th}, \& {Gombosi}}]{van-der-Holst:2014aa}
{van der Holst}, B., {Sokolov}, I.~V., {Meng}, X., {et~al.} 2014, \apj, 782, 81

\bibitem[{{Velli} {et~al.}(2015){Velli}, {Pucci}, {Rappazzo}, \&
  {Tenerani}}]{Velli:2015aa}
{Velli}, M., {Pucci}, F., {Rappazzo}, F., \& {Tenerani}, A. 2015, Philosophical
  Transactions of the Royal Society of London Series A, 373, 20140262

\bibitem[{{Vidotto} {et~al.}(2014{\natexlab{a}}){Vidotto}, {Jardine}, {Morin},
  {Donati}, {Opher}, \& {Gombosi}}]{Vidotto:2014aa}
{Vidotto}, A.~A., {Jardine}, M., {Morin}, J., {et~al.} 2014{\natexlab{a}},
  \mnras, 438, 1162

\bibitem[{{Vidotto} {et~al.}(2009){Vidotto}, {Opher}, {Jatenco-Pereira}, \&
  {Gombosi}}]{Vidotto:2009ab}
{Vidotto}, A.~A., {Opher}, M., {Jatenco-Pereira}, V., \& {Gombosi}, T.~I. 2009,
  \apj, 699, 441

\bibitem[{{Vidotto} {et~al.}(2014{\natexlab{b}}){Vidotto}, {Gregory},
  {Jardine}, {Donati}, {Petit}, {Morin}, {Folsom}, {Bouvier}, {Cameron},
  {Hussain}, {Marsden}, {Waite}, {Fares}, {Jeffers}, \& {do
  Nascimento}}]{Vidotto:2014ab}
{Vidotto}, A.~A., {Gregory}, S.~G., {Jardine}, M., {et~al.} 2014{\natexlab{b}},
  \mnras, 441, 2361

\bibitem[{{Wang} \& {Sheeley}(1991)}]{Wang:1991aa}
{Wang}, Y.-M., \& {Sheeley}, Jr., N.~R. 1991, \apjl, 372, L45

\bibitem[{{Washimi} \& {Shibata}(1993)}]{Washimi:1993aa}
{Washimi}, H., \& {Shibata}, S. 1993, \mnras, 262, 936

\bibitem[{{Weber} \& {Davis}(1967)}]{Weber:1967aa}
{Weber}, E.~J., \& {Davis}, Jr., L. 1967, \apj, 148, 217

\bibitem[{{Wood} {et~al.}(2014){Wood}, {M{\"u}ller}, {Redfield}, \&
  {Edelman}}]{Wood:2014aa}
{Wood}, B.~E., {M{\"u}ller}, H.-R., {Redfield}, S., \& {Edelman}, E. 2014,
  \apjl, 781, L33

\bibitem[{{Wood} {et~al.}(2002){Wood}, {M{\"u}ller}, {Zank}, \&
  {Linsky}}]{Wood:2002aa}
{Wood}, B.~E., {M{\"u}ller}, H.-R., {Zank}, G.~P., \& {Linsky}, J.~L. 2002,
  \apj, 574, 412

\bibitem[{{Wright} {et~al.}(2011){Wright}, {Drake}, {Mamajek}, \&
  {Henry}}]{Wright:2011aa}
{Wright}, N.~J., {Drake}, J.~J., {Mamajek}, E.~E., \& {Henry}, G.~W. 2011,
  \apj, 743, 48

\bibitem[{{Zanni} \& {Ferreira}(2009)}]{Zanni:2009aa}
{Zanni}, C., \& {Ferreira}, J. 2009, \aap, 508, 1117

\end{thebibliography}
\end{document}